\documentclass[numberedappendix]{emulateapj}
\newcommand{\Mb}{M_\mathrm{B}}
\newcommand{\mb}{m_\mathrm{B}}
\newcommand{\mr}{m_\mathrm{r}}

\newcommand{\smap}{a_\mathrm{p}}

\slugcomment{Accepted by The Astrophysical Journal}
\shorttitle{Virgo early-type dwarfs. III. Subpopulations}
\shortauthors{Lisker, Grebel, Binggeli, \& Glatt}
\begin{document}
 
\title{Virgo cluster early-type dwarf galaxies with the Sloan Digital
  Sky Survey.\\
III. Subpopulations: distributions, shapes, origins}

\author{Thorsten Lisker, Eva
  K. Grebel, Bruno Binggeli, and Katharina Glatt}
\affil{Astronomical Institute, Dept.\ of Physics and Astronomy,
  University of Basel, Venusstrasse 7, CH-4102 Binningen, Switzerland}
\email{tlisker@astro.unibas.ch,grebel@astro.unibas.ch,\\binggeli@astro.unibas.ch,glatt@astro.unibas.ch}

\begin{abstract}
From a quantitative analysis of 413 Virgo cluster early-type dwarf galaxies
(dEs) with Sloan Digital Sky Survey imaging data, we find that the dE
class can be divided into multiple subpopulations that differ
significantly in their morphology and clustering properties. Three dE
subclasses are shaped like thick disks and show no central clustering:
(1) dEs with disk features like spiral arms or bars, (2) dEs with 
central star formation, and (3) ordinary, bright dEs that have no or
only a weak nucleus. These populations probably formed from infalling
progenitor galaxies. In contrast, ordinary nucleated dEs follow the
picture of classical dwarf elliptical galaxies in that they are
spheroidal objects and are centrally clustered like E and S0 galaxies,
indicating that they have resided in the cluster since a long time, or
were formed along with it. These results define a morphology-density
relation \emph{within} the dE class.
We find that the difference in the clustering properties of nucleated dEs 
and dEs with no or only a weak nucleus is not caused by selection biases, 
as opposed to previously reported suggestions.
The correlation between surface brightness and observed axial ratio favors
oblate shapes for all subclasses, but our derivation of
intrinsic axial ratios indicates the presence of at least some triaxiality.
 We discuss possible interrelations and formation 
mechanisms (ram-pressure stripping, tidally induced star formation,
harassment) of these dE subpopulations.
\end{abstract}
 
 
%________________________________________________________________

\section{Introduction}
 \label{sec:intro}

As the most numerous type of galaxy in clusters, early-type dwarf
galaxies are ideal probes to study the physical processes that govern
galaxy formation and evolution in environments of different density. 
The pronounced morphology-density relation
\citep[e.g.,][]{dre80,bin87} suggests that early-type dwarfs were either
formed mainly in high-density environments, or 
originate from galaxies that fell into a cluster and were
morphologically transformed. However, the actual formation mechanisms
are still a matter of debate \citep[see][and
  references therein]{iauc198}. Most of the proposed
scenarios are based on the vigorous
forces acting within a cluster environment, like ram-pressure stripping
\citep{gun72} of dwarf irregular (dIrr) galaxies
\citep[e.g.,][]{vZe04a}, tidally induced star formation in dIrrs
\citep{dav88}, or so-called harassment \citep{moo96} of
infalling late-type spirals through close encounters with massive
cluster members. 

Early-type dwarfs 
form a rather heterogeneous class of objects. In addition to the
classical dwarf ellipticals,
\citet{san84} introduced the class of dwarf S0 (dS0) 
galaxies, which were conjectured to have
disk components, based on signatures like high flattening or a
bulge+disk-like profile \citep{bin91}.
The identification of spiral 
substructure then provided the first direct proof for a disk in an
early-type dwarf \citep{jer00a}, which, however, had not been
classified as dS0 but as dwarf elliptical. Inspired by similar discoveries
\citep[e.g.,][]{bar02a,gra03,der03}, we performed a search for disk
features in 410 Virgo cluster early-type dwarfs \citep[hereafter Paper I]{p1}. We thereby included
galaxies classified as dwarf elliptical and as dS0 to avoid any
preselection bias, and assigned them the common abbreviation
``dE'', which we adopt for this Paper as well. We identified disk features in 36 dEs,
 and argued that they constitute an unrelaxed population of
 disk-shaped galaxies different from the classical dwarf ellipticals
 (Paper I).

But the dE class shows yet more diversity: nucleated and
non-nucleated dEs have different clustering properties
\citep{vdB86,fer89}, their flattening distributions differ
(\citealt{bin91}; \citealt{ryd94,bin95}),
%\citep{bin91,ryd94,bin95},
and color differences were reported as well
\citep{rak04,lis05}. Moreover, several
of the bright dEs display blue central regions caused by recent or ongoing
star formation \citep[hereafter Paper II]{p2}, and also differ in their spatial and flattening
distributions from the bulk of dEs. Thus, prior to discussing possible
formation mechanisms, we need to
systematically disentangle the various dE subclasses
observationally. This is the purpose of this Paper.

%________________________________________________________________

\section{Sample selection}
\label{sec:sample}

While our dE sample selection was already described in Papers I and II
of this series, these studies were still based on the Data Release 4 of
the Sloan Digital Sky Survey \citep[SDSS;][]{sdssdr4}. Since we are now using
the full SDSS Data Release 5 \citep[DR5,][]{sdssdr5} dataset, we
provide here a detailed, updated description of our selection.

\subsection{Selection process}
\label{sec:sub_select}

The Virgo Cluster Catalog \citep[VCC,][]{vcc}, along with revised
classifications from \citet[VCC 1422]{bar02a}, \citet[VCC 0850]{bar03},
and \citet[VCC 1488]{geh03}, contains 1197 galaxies classified ``dE''
or ``dS0'', including candidates, that are certain or possible
cluster members according to \citet{vcc}, \citet{virgokin}, and Paper
II. 552 of these fall within our chosen limit in apparent B magnitude
from the VCC of $\mb\le18.0$ mag (see Paper I). This is the same magnitude
limit up to which the VCC was found to be complete \citep{vcc}.
When adopting a Virgo cluster distance of $d=15.85 \rm{Mpc}$, i.e., a
distance modulus $m-M=31.0$ mag \citep[see, e.g.,][]{fer00}, which we use
throughout, this corresponds roughly to a limit in
  absolute magnitude of $\Mb\le -13.0$ mag.
 
Six galaxies are not
covered by the SDSS.
 While we initially included objects with uncertain classification
 (e.g., ``dE?''),
 we then excluded all 50 galaxies that appeared to
be possible dwarf irregulars from visual inspection of the coadded
SDSS $g$,r, and $i$ images (see Paper I), or were classified as ``dE/Im''.
 Three more objects (VCC 0184, 0211, and 1941) were excluded because they
 appear to be probable background spirals.
Finally, VCC 1667 could not be classified properly, since it is
significantly blended with multiple other galaxies. This leads to a
final dE sample of 492 certain or possible cluster members, containing
426 certain cluster members on which we focus in the present Paper.

\subsection{Presence of nuclei}
\label{sec:sub_nuc}

\begin{figure}
  \epsscale{0.5}
  \plotone{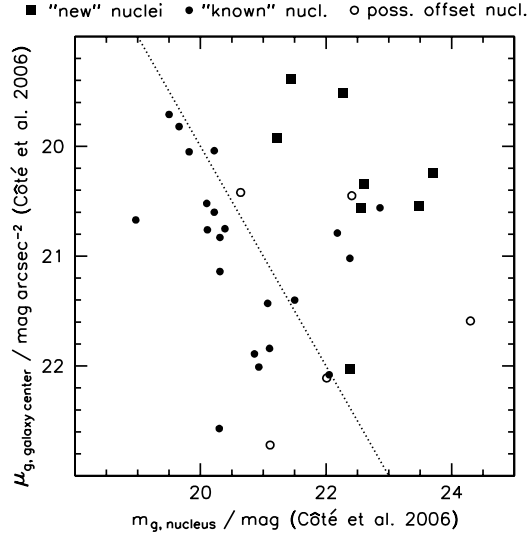}
  \caption{Nucleus detection limits.
      Shown are central surface brightnesses and nucleus magnitudes in
      $g$, both from \citet{acsvcs8}, for 34 of the 36 VCC dEs of their
      sample (VCC 1512 has no nucleus, and for VCC 1743 no values could
      be derived). Objects where \citet{acsvcs8} identified ``new''
      nuclei, i.e., that are listed as non-nucleated or only possibly
      nucleated in the VCC, are shown as filled squares. Objects that
      might have an offset nucleus according to \citet{acsvcs8}
      and that are listed as non-nucleated in the VCC are shown as open
      circles. Objects classified as nucleated in both \citet{acsvcs8}
      and the VCC (including class ``N:'') are represented by the
      filled circles. The dotted line follows equal values
      of central surface brightness (in
      mag arcsec$^{-2}$) and nucleus magnitude (in mag).
  }
  \label{fig:nuc}
\end{figure}

While our classification of nucleated and non-nucleated
dEs relies on
the VCC,
it is known from HST
observations that many apparently non-nucleated dEs actually host a
faint nucleus hardly detectable with ground-based imaging
\citep[also see \citealt{lotz04}]{acsvcs8}.
A direct comparison of the VCC classification with the results from
\citet{acsvcs8} shows that, as a rough rule of thumb, the detection of dE nuclei
in the VCC becomes incomplete for nucleus magnitudes
that are fainter than the respective value of the host
galaxy's central surface brightness, measured within a radius $r=0.1''$
(Fig.~\ref{fig:nuc}).
Our non-nucleated dEs could thus be more appropriately termed
\emph{dEs without a nucleus of significant relative brightness}
as compared to the underlying light of the galaxy's center. 
In fact, \citet{gra05}
suggested that dEs classified as nucleated and non-nucleated might
actually form
a continuum of dEs with respect to relative nucleus
brightness. Therefore, the VCC 
classification basically translates into probing opposite sides of
this continuum --- and this is exactly what makes it useful for
our study of dE subclasses. If the relative
brightness of a nucleus depends on its host galaxy's evolutionary
history, then one might expect nucleated and ``VCC-non-nucleated'' dEs
to exhibit different population properties.

\section{Data}
\label{sec:data}

  The SDSS DR5 covers
  all VCC galaxies except for an
  approximately $2\arcdeg \times 2\fdg5$ area at $\alpha\approx 186\fdg2$,
  $\delta\approx +5\fdg0$. It provides reduced images taken in the 
  $u$, $g$, $r$, $i$, and $z$ bands with an effective exposure time of
  $54 \rm{s}$ in each band \citep[see also][]{sdssedr}, as well as the
  necessary parameters to flux calibrate them. The pixel scale
  of $0\farcs396$ 
  corresponds to a physical size of $30$ pc at our adopted Virgo
  cluster distance of $d=15.85 \rm{Mpc}$. The SDSS imaging camera
  \citep{sdsscamera} takes data in drift-scanning 
  mode nearly simultaneously in the five photometric bands, and thus
  combines very homogeneous multicolor photometry 
  with large area coverage and sufficient depth to
  enable a systematic analysis of dEs.
  The images have an absolute astrometric accuracy of $rms \le
  0\farcs1$ per coordinate, and a relative accuracy between the $r$ band
  and each of the other bands of less than $0.1$ pixels
  \citep{sdssastrometry}. They can thus easily be aligned using
  their astrometric calibration and need not be registered
  manually.

  The $rms$ of the noise per pixel corresponds to a
  surface brightness of approximately $24.2$ mag arcsec$^{-2}$ in the
  u-band, $24.7$ in $g$, $24.4$ in $r$, $23.9$ in $i$, and $22.4$ in z.
  The typical total signal-to-noise ratio 
  (S/N) of a bright dE ($m_\mathrm{B,VCC}\approx 14$) amounts to about $1000$
  in the r-band within an aperture radius of
  approximately two half-light radii. For a faint dE ($\mb \approx 18$) this
  value is typically about $50$. While the S/N in the $g$ and i-band is
  similar, it is several times lower in the z-band
  and more than ten times lower in the u-band.

The SDSS provides photometric measurements for our galaxies, but
we found these to be incorrect in many cases \citep{lis05}. The SDSS
photometric pipeline significantly overestimates the local sky flux
around the Virgo dEs due to their large apparent sizes and low surface
brightness outskirts. This affects the derivation of isophotal and
Petrosian radii, the profile fits, and subsequently the calculation of
total magnitudes, which can be wrong by up to 0.5 mag.
For this reason, we used $B$ magnitudes from the VCC throughout the
first two papers of this series. In the meantime, we have performed our own
structural and photometric measurements (see Sect.~\ref{sec:image}),
which we shall use here as well as in future papers of this
series.
Still, when we refer to $B$ magnitudes, these were adopted from
  the VCC.

Heliocentric velocities for part of the sample are provided by the
NASA/IPAC Extragalactic Database (NED;
also see Paper II for more detailed references).

\section{Image preparation and analysis}
\label{sec:image}

\subsection{Sky subtraction}
\label{sec:sub_sky}

The sky level on the SDSS images can vary by some tenths of the noise
level across an image. For a proper determination of Petrosian radii
 of the dEs (see Sect.~\ref{sec:sub_morpho}) despite their low surface
brightness outskirts,
it is thus not always sufficient to subtract only a single sky flux
value from each SDSS image. Therefore, we performed sky subtraction through
the following procedure. First, we constructed object masks for each
SDSS image from the
so-called segmentation images of the Source Extractor software
\citep{sex} by expanding these through smoothing with a Gaussian filter (using
IRAF\footnote{IRAF is distributed by the National Optical
  Astronomy Observatories, which are operated by the Association of
  Universities for Research in Astronomy, Inc., under cooperative
  agreement with the National Science Foundation.},
\citealt{iraf}). A preliminary sky level was then determined for each
image as the median of all unmasked pixels, clipped three times
iteratively at $3
\sigma$. In order to reach a higher S/N than that of the individual images, we
  then produced a coadded image by summing the (weighted) $g$, $r$, and
  i-band images as described in Paper I. We then obtained an improved
  object mask from the coadded image and used this to refine our sky level
  measurement.

Finally, the sky flux distribution across the image was determined by
  computing the average flux -- clipped five times iteratively at $3 \sigma$ -- of
  all unmasked pixels in 201$\times$201 pixel boxes, centered every 40
  pixels. This grid of values can be stored as a 52$\times$38 pixel
  ``sky image''. Pixels in this sky image that did not contain useful values due to
  too many masked pixels in the parent image were linearly interpolated using IRAF
  \emph{fixpix}. We then applied a 3$\times$3 pixel median filter to
  the sky image, expanded it to match its parent SDSS image's size
  (using IRAF \emph{magnify} with linear interpolation), and
  subtracted it from the latter. This yields the final $u$, $g$, $r$, $i$, and
  $z$ images.

We point out that there is, to our knowledge, no general agreement or recipe
as to whether to use, 
e.g., the clipped mean, the median, the clipped median, or the mode,
for determination of the sky level. However, it is advisable that
the chosen approach be reconciled with the image measurements to be
performed, which in our case is the derivation of Petrosian radii (see
Sect.~\ref{sec:sub_morpho}). Since the latter is based on the \emph{average}
flux within given annuli, we chose to use the clipped \emph{average}
flux of all unmasked pixels for our sky level measurement. This
guarantees that the resulting flux level in each image is zero as
``seen'' by the Petrosian radius calculation.\footnote{
The reason why such considerations are at all necessary is the same as that
for which the SDSS pipeline overestimated the local sky flux: the Virgo
dEs are large in apparent size and cover 10$^{4}$ to
10$^{5}$ pixels, but their low surface brightness outskirts cause a
large number of these pixels to have S/N$<1$. Thus, a wrong sky level
estimate of the order of just a few tenths of the noise level can have
a large effect in total.
}

\subsection{Calibration and extraction}
\label{sec:sub_extract}

We calibrated the sky subtracted SDSS images using the provided flux
calibration information (photometric zeropoint and airmass
correction). We also corrected for the reported SDSS zeropoint offsets
in the $u$ and $z$ bands from
the AB system \citep[see
  http://www.sdss.org/dr5/algorithms/fluxcal.html]{ABsystem}.
 However, before working with the images, it is
advantageous to put together adjacent images: a number of
galaxies partly extend beyond the image edges and reappear on the
corresponding neighbouring image. Bright dEs typically
have apparent diameters of 300 pixels or more, which is rather large
compared to the SDSS image size of 2048$\times$1489 pixels. The SDSS
astrometric calibration allows us to accurately put together adjacent
images, which we did before extracting an 801$\times$801 pixel cutout
image for each galaxy. These cutout images were then corrected for
Galactic extinction, using one value per image, calculated with the
dust maps and corresponding software of \citet[provided at
  http://www.astro.princeton.edu/$\sim$schlegel/dust/data/]{sch98}.
From the $g$, $r$, and $i$ cutout image we produced a final coadded image for
each galaxy.

\subsection{Morphology}
\label{sec:sub_morpho}

We perform an iterative process of determining shape and total flux
for each galaxy, as described below. Throughout this process, we mask
disturbing foreground or background objects, i.e., we do not consider
masked pixels in any calculation.
We start with deriving the Petrosian radius \citep{pet76}, as defined by
\citet{sdssedr}, on the coadded image. Using a circular aperture
with one Petrosian radius, we then find the center of the galaxy's image by
iteratively searching for the minimum asymmetry, following
\citet{con00}. The asymmetry $A$ is calculated as
\begin{equation}
A = \frac{\displaystyle\sum_i\vert f_i-f_{i,180}\vert}{\displaystyle\sum_i\vert f_i\vert},
\end{equation}
where $f_i$ is the flux value of the i-th pixel, and $f_{i,180}$ is the
flux value of the corresponding pixel in the 180-degree rotated image.

 The asymmetry is computed using an initially guessed
central position (from Paper I for objects in the SDSS DR4, and from visual
examination for objects in DR5, using SAOImage DS9, \citealt{ds9}), as
well as for using the surrounding eight positions in a 3$\times$3 grid
as center. If one of the surrounding positions yields a lower 
asymmetry, it is adopted as new central position. This process is
repeated until convergence. We perform two of these ``asymmetry
centerings'': a first one with a step size of 1 pixel, and a second one
with a step size of 0.3 pixels. The initial and final value typically
differ by less than a pixel.

We then compute the parameters defining an elliptical aperture (axial
ratio and position angle) from the image moments \citep{abr94}, and derive
a ``Petrosian semimajor axis'' $\smap$, i.e., we use ellipses instead of
circles in the calculation of the Petrosian radius \citep[see, e.g.,][]{lot04}.
Within this elliptical aperture with a semimajor axis
$a$ of $1 \smap$, we perform another iteration to re-derive the
elliptical shape parameters from the image moments, and also to re-derive
 $\smap$.

The elliptical shape is then applied to measure the total
flux in the $r$ band within an elliptical aperture with $a = 2
\smap$, which also yields a half-light semimajor axis in $r$ ($a_{{\rm hl},r}$). Using
this value for $a_{{\rm hl},r}$, we go back to the coadded image and
fit an ellipse to
the isophotal shape of the galaxy at $a = 2 a_{{\rm hl},r}$, using
IRAF \emph{ellipse}. The elliptical annulus used for the isophotal fit
ranges from
$2^{0.75}a_{{\rm hl},r}$ to $2^{1.25}a_{{\rm hl},r}$.

This new elliptical shape is now used to derive the final Petrosian
semimajor axis on the coadded image, and to subsequently measure again
the total flux in the $r$ band within $a = 2 \smap$, yielding the final
value for $a_{{\rm hl},r}$. The isophotal shape is then measured
  again at $a = 2 a_{{\rm hl},r}$, yielding the axial ratio that we
  shall use throughout this Paper.

Since we masked disturbing foreground or background objects by not considering
their pixels, our measured total flux for a given galaxy is always lower than
it would be without any such ``holes'' in the galaxy's image. In order to
correct for this effect, we subdivide the final aperture of each galaxy into 20
elliptical annuli of equal width, and assign each masked pixel the
average flux value of its respective annulus. This yields our final value for
the total $r$ band flux and the corresponding magnitude. The difference to the
uncorrected value is typically less than 0.1 mag.

For 13 of our dEs, the derivation of the Petrosian radius did not
converge, due to the fact that these galaxies sit within the light of
nearby bright sources. While in some of these cases, it would still be
possible to ``manually'' define an axial ratio for the galaxy, we
decided to exclude these objects from our sample, since no reliable
$r$ band magnitudes can be derived, which are needed for our
definitions of dE subclasses in Sect.~\ref{sec:subclasses}. This leaves
us with a working sample of 413 Virgo cluster dEs.

%________________________________________________________________

\section{Early-type dwarf subclasses}
\label{sec:subclasses}

\subsection{Subclass definitions}
\label{sec:sub_subclass}

%pics_normal:
%0929,0750,1573
%
%pic_dE(di):
%0308
%
%pic_dE(bc):
%0021

\begin{figure}
  \epsscale{0.8}
  \plotone{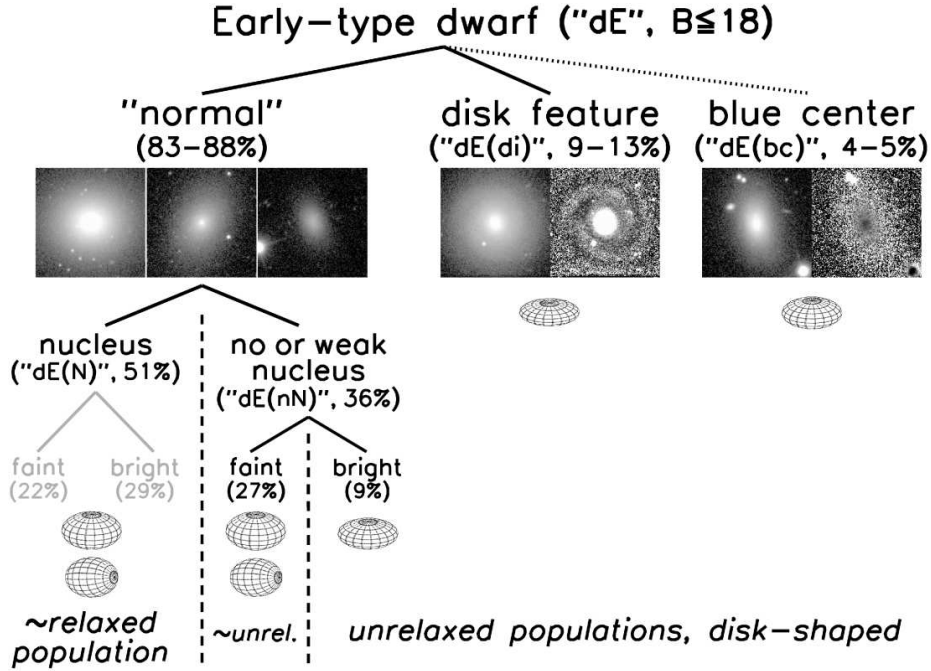}
  \caption{The zoo of early-type dwarfs. Our subdivision is
  shown as tree scheme, along with a 3-D illustration of a typical
  intrinsic shape, using the median axial ratio from the intrinsic
  distributions shown in Fig.~\ref{fig:multiI} (see
  Sect.~\ref{sec:sub_shape}). We also provide a statement about
  the inferred dynamical status of each subpopulation, as indicated by
  their clustering properties: ``relaxed'' and ``unrelaxed'' (see Sect.~\ref{sec:sub_cluster}). The subdivision of dE(N)s into faint and
  bright samples is shown in grey color to reflect the fact that we
  find them to be not different in their properties (see Sect.~\ref{sec:subclasses}). The branch
  of the dE(bc)s is shown as dotted line only, since these are not a
  \emph{morphological} subclass (see text). The
  percentage of each subsample among our 413 Virgo cluster dEs is given
  in parentheses. For the dE(bc)s and dE(di)s 
  the percentage ranges include corrections for the estimated number of
  objects missed by our detection techniques (see Papers I and II).
  Three sample images are shown for the normal dEs. For the dE(di)s, we
  show one sample image along with its unsharp mask revealing the
  spiral substructure. For the dE(bc)s, one sample image is shown along
  with its $g-i$ color map revealing the blue center (dark=blue). See
  text for further details.
  }
  \label{fig:tree}
\end{figure}

Of our 413 Virgo dEs, 37 display disk features, like spiral
arms, bars, or signs of an edge-on disk (Paper I,
adding VCC 0751 to the objects listed there in order to update to SDSS
DR5). We term these objects ``dE(di)s'', and separate this dE subclass
from the ordinary, ``featureless'' dEs (Fig.~\ref{fig:tree}).
In order to further explore the diversity of the latter, we
perform a secondary subdivision into nucleated (``dE(N)'') and
non-nucleated (``dE(nN)'') galaxies, based on the identification of
nuclei in the VCC as outlined in
Sect.~\ref{sec:sub_nuc}. Since a further subdivision of the dE(di)s
would lead to statistically insignificant subsamples, we shall instead
discuss their nucleated fraction in the text. Finally, since our
galaxies span a range of almost 5 mag in $r$, it appears worth
performing a tertiary subdivision into dEs brighter and fainter than
the median $r$ brightness of our full sample, namely $\mr=15.67$
mag. Moreover, all but three of the dE(di)s are brighter than this
value; thus our subdivision allows us to compare them to ordinary dEs of
similar luminosities. The percentage of each subsample among our full
sample of 413 dEs is given in parentheses in Fig.~\ref{fig:tree},
whereas the actual number of galaxies contained in each subsample
is given in the left column of Fig.~\ref{fig:multiI}.

The subclasses defined so far are based on structural properties only
--- for \emph{morphological} classification of galaxies, it is not
advisable to use color information. However, in Paper II we identified
a significant number of dEs with blue centers
(17 galaxies, including VCC 0901 from the SDSS DR5).
These objects, termed ``dE(bc)s'', exhibit recent or ongoing central star
formation, similar to NGC 205 in the Local Group.
They were morphologically classified as dwarf
ellipticals or dS0s by \citet{san84}, and their regular, early-type
morphology was confirmed in Paper II; thus, they are not possible
irregular galaxies, which we have excluded from our samples here and in
previous papers of this series.
 The flattening distribution of the dE(bc)s was found to be incompatible with
 intrinsically spheroidal objects (Paper II), and their 
distribution with respect to local projected density suggests that they are an
unrelaxed population. The latter result is similar to the spatial
distribution of Virgo and Fornax dwarfs with early-type morphology that
are gas-rich and/or show star formation \citep{dri01,conIV,buy05}.

While it is not clear a priori that any of the dE 
subclasses defined above are evolutionary interrelated, each dE(bc)
unavoidably evolves into one of the above dE types once star formation
ceases and the central color reddens (Paper II). Therefore, and because
the dE(bc)s are defined through color instead of morphological
properties, we do not consider them a \emph{morphological} dE
subclass.\footnote{
 A \emph{morphological} peculiarity of several dE(bc)s is that
 they show central irregularities, which are presumably due to
 gas, dust, and/or star formation, similar to NGC
 205. These can be seen, e.g., when constructing unsharp mask images (Paper
 II). However, an attempt to quantify these weak features through image
 parameters like asymmetry or clumpiness yielded no clear separation
 from the bulk of dEs. Moreover, not all dE(bc)s display such features.
}
On the other hand, their star formation and presence of gas
(Paper II) might imply that their formation process is not completely
finished yet. It thus appears more cautious to separate them from the
rest of dEs (see Fig.~\ref{fig:tree}) in order to not bias the
population properties of the other subclasses. In the discussion
(Sect.~\ref{sec:discussion}) we try to assess which dE type(s) the
dE(bc)s could possibly evolve into. Note that four objects are common
to both the dE(di) and the dE(bc) sample. We exclude these from
the sample of dE(di)s, which now comprises 33
galaxies. Table~\ref{tab:zoo} lists our dEs along with their subclass.

A similar subdivision of the dE class into bright and faint
(non-)nucleated subsamples was performed by \citet{fer89}, also with
the aim of studying shapes and spatial distributions of the resulting
subsamples. Our subdivision is different in two respects: first,
\citeauthor{fer89} defined all galaxies with $\mb< 17.5$ mag as ``bright'',
whereas our magnitude separation (at $\mr= 15.67$ mag) is done at
significantly brighter values and divides our full sample into equally
sized halves. Second, we have the advantage of excluding dE(di)s and
dE(bc)s from the ``normal'' dEs, thereby obtaining cleaner subsamples,
especially for the bright objects: all but three of the dE(di)s are
brighter than $\mr= 15.67$ mag.

While \citet{fer89} found statistically significant differences in the
 spatial distributions of their subsamples -- with dE(N)s being much
 more centrally clustered than the bright dE(nN)s -- their flattening
 distributions were only based on eye-estimated axial ratios from
 photographic plates. These can be uncertain by $\sim$20\%
 \citep{fer89}. With our
 measured axial ratios from the coadded SDSS images at hand, we
 therefore present in the following subsection a more detailed and
 accurate study of the flattening distributions of the different dE
 subsamples, and attempt to deduce their approximate intrinsic shapes.

\subsection{Subclass shapes}
\label{sec:sub_shape}

\begin{figure}
  \epsscale{1.0}
  \plotone{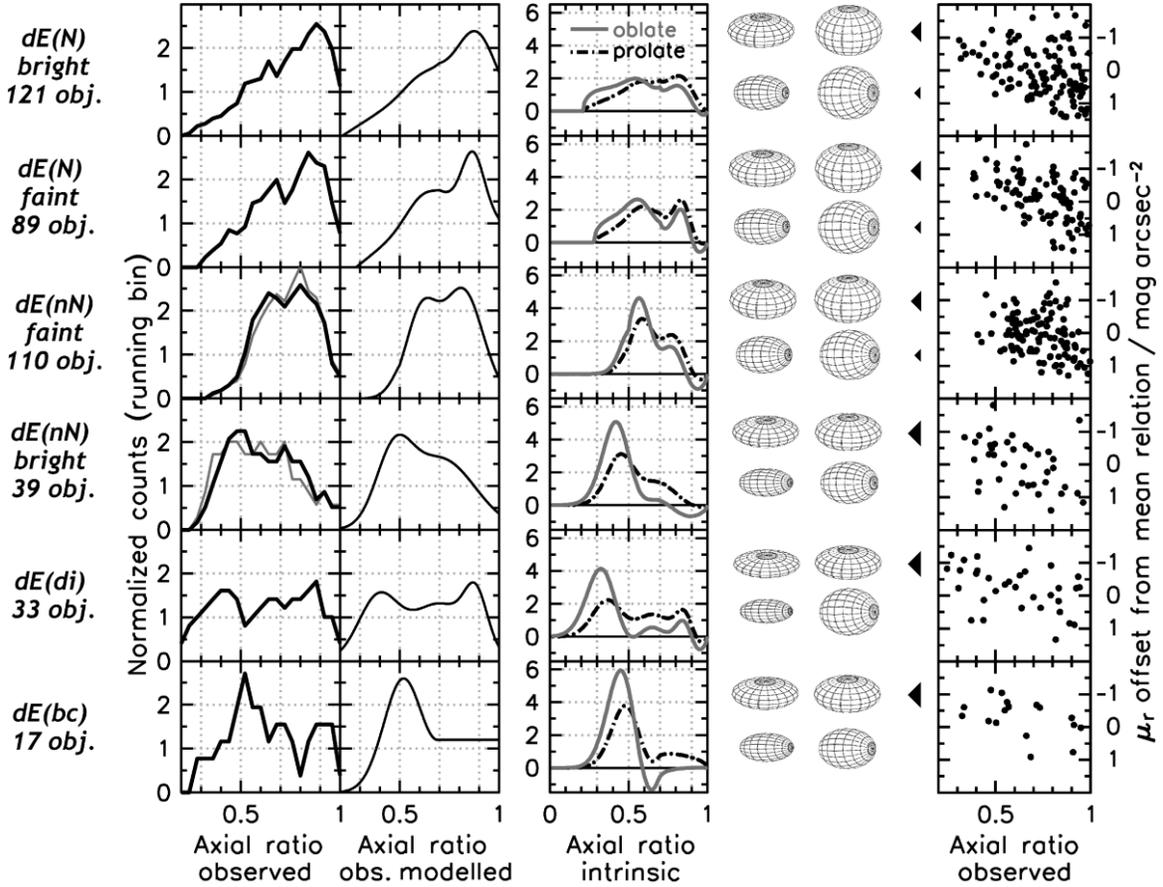}
  \caption{Deduction of intrinsic shapes.
      In the leftmost column, the number of objects in each of our dE
      subsamples are given. 
      In the second column, we show the distribution of projected axial
      ratios for each of our dE subsamples.
      The data are shown as running
      histogram with a bin width of 0.15, corresponding to one fifth
      of the range in axial ratio covered by our galaxies. Each curve
      is sampled in steps of 0.04 (one quarter of the bin width) and
      normalized to an area of 1.
      The observed distributions are approximated by analytic
      functions (see text), shown in the third column. From these, we
      derive intrinsic axial ratio distributions (fourth column),
      adopting purely oblate (grey solid line) and purely prolate
      (black dash-dotted line) shapes. For these distributions, we show
      in the fifth column 3-D illustrations of the galaxy shapes for
      the oblate (upper) and prolate (lower) case. For
      each distribution, we show the shape when using the 25th
      percentile axial ratio (left) and the 75th percentile axial
      ratio (right). In the rightmost column, we show for each dE
      subsample the surface brightness test (see text): we plot the
      surface brightness offset from the mean relation of $r$ band
      surface brightness and magnitude against axial
      ratio. Surface brightness is measured within $a = 2 a_{{\rm hl},r}$,
      since axial ratio is measured at the same semimajor axis. The mean
      relation of surface brightness and magnitude is obtained through
      a linear least squares fit with one 3$\sigma$-clipping.
      The arrows pointing from the surface brightness test diagrams
      towards the shape illustrations reflect whether the test implies
      oblate or prolate intrinsic shapes; see text for more details.
  }
  \label{fig:multiI}
\end{figure}

From the axial ratio measurements of our galaxies
(Sect.~\ref{sec:sub_morpho}), we put together the flattening
distributions of each dE subsample. These are presented in the second
column of Fig.~\ref{fig:multiI} as running histograms, i.e., at each
sampling point we consider the number of objects within a bin of constant
width, and normalize the resulting curve to an area of 1. The bin width
is 0.15, which we have chosen to be one fifth of the range in axial
ratio covered by our galaxies. The sampling step is 0.04 (one quarter
of the bin width). The bright and faint dE(N)s, and also the faint
dE(nN)s, predominantly have rather round apparent shapes, while the
bright dE(nN)s, dE(di)s, and dE(bc)s exhibit a significant fraction of
objects with rather flat apparent shapes.

Since the division between bright and faint objects at $\mr=15.67$ mag
is somewhat arbitrary, we test whether the difference between the axial
ratio distributions of faint and bright dE(nN)s becomes even more
pronounced if a wider magnitude separation is adopted. The grey curves
in the respective panels of the second column of Fig.~\ref{fig:multiI}
show the distributions for bright dE(nN)s with $\mr\le15.67-0.5$ mag
(23 objects) and for faint dE(nN)s with  $\mr\le15.67+0.5$ mag (86
objects). While the faint dE(nN)s basically remain unchanged, the
bright dE(nN)s indeed tend slightly towards flatter shapes, but the
difference is rather small.

\begin{figure}
  \epsscale{0.5}
  \plotone{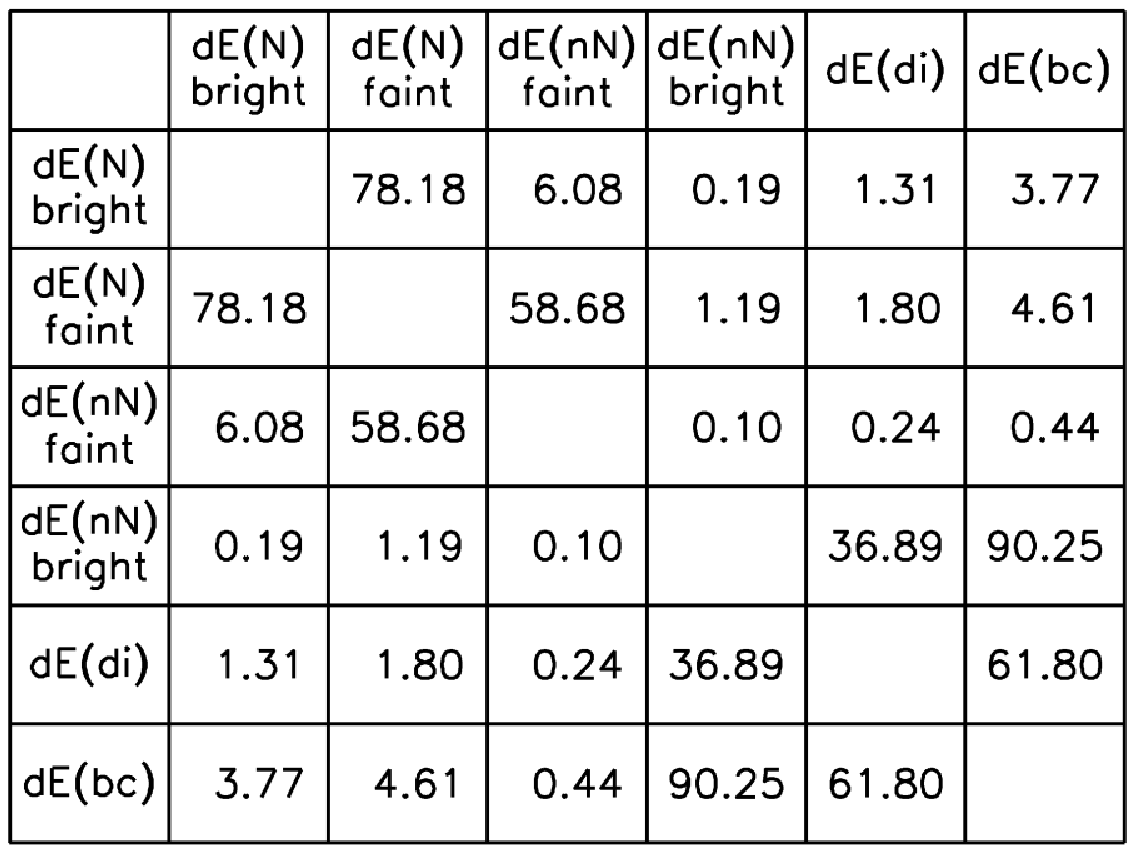}
  \caption{
K-S test results for the comparison of the axial ratio distributions of our
dE subsamples. For each pair of distributions we give the probability
in percent for the null hypothesis that the two distributions stem from the same
underlying distribution function.
  }
  \label{fig:ax_ks}
\end{figure}

A statistical comparison of the axial ratio distributions of our dE
subsamples confirms what is seen in Fig.~\ref{fig:multiI}: a K-S test
yields very low probabilities that any of the ``flatter'' subsamples
(bright dE(nN)s, dE(di)s, and dE(bc)s, lower three rows) could stem from the same true
distribution function as any of the ``rounder'' subsamples (bright and
faint dE(N)s as well as faint dE(nN)s, upper three rows). This confirms our findings from
Papers I and II for the dE(di)s and dE(bc)s, respectively. The
resulting probabilities from the K-S test for the pairwise comparison
of the subsamples are given as percentages in Fig.~\ref{fig:ax_ks}. Interestingly, the
lowest probability of all comparisons is obtained when matching the
distributions of bright and faint dE(nN)s: here, the probability of the
null hypothesis that they stem from the same underlying distribution
function is only 0.10\%. Note that the probabilities for the comparison
of the ``flatter'' subsamples with the ``rounder'' ones increase
slightly with decreasing sample size, going from the bright dE(nN)s to
the dE(di)s and then to the dE(bc)s. However, the probability for a
common underlying distribution of dE(bc)s and the bright and faint
dE(N)s is still only 3.8\% and 4.6\%, respectively.

Is it possible to deduce the distributions of \emph{intrinsic} axial
ratios from those of the apparent ones? As discussed in detail by
\citet{bin95}, the intrinsic shapes can be deduced when assuming that
they are purely oblate or purely prolate. The distribution function
$\Psi$ of intrinsic axial ratios $q$ can then be derived from the
distribution function $\Phi$ of observed axial ratios $p$ through (\citealt{fal83},
eqs.\ (6) and (9))
\begin{equation}
\Psi(q) = \frac{2}{\pi} \sqrt{1-q^2} \frac{d}{dq} \int_0^q dp \frac{\Phi(p)}{\sqrt{q^2-p^2}}
\end{equation}
for the oblate case, and
\begin{equation}
\Psi(q) = \frac{2}{\pi} \frac{\sqrt{1-q^2}}{q^2} \frac{d}{dq} \int_0^q dp \frac{p^3\Phi(p)}{\sqrt{q^2-p^2}}
\end{equation}
for the prolate case. Following \citet{bin95}, we first defined
adequate analytic functions $\Phi(p)$ that represent the
observed distributions, and then evaluated the above equations
numerically. The analytic ``model functions'' are shown in the third
column of Fig.~\ref{fig:multiI}; they were constructed from
combinations of (skewed) Gaussians with each other and, in some cases,
with straight lines. Note that, for the dE(bc)s, we decided not to
follow the observed distribution in all detail, since it is drawn from
a rather small sample of 17 galaxies, which probably is the cause of
the fluctuations seen.

The deduced intrinsic distributions are presented in the fourth column
of Fig.~\ref{fig:multiI}, for the oblate (grey lines) and prolate
(black dash-dotted lines) case. We also show 3-D illustrations of the
galaxy shapes for each distribution (fifth column), using in each case
the axial ratio of the 25th percentile (left 3-D plot) and the 75th
percentile (right 3-D plot).
These results confirm that the bright dE(nN)s, dE(di)s, and dE(bc)s do
have lower axial ratios than the bright and faint dE(N)s and the faint
dE(nN)s. Furthermore, we point out that 
the bright and faint dE(N)s span a rather wide range of
intrinsic axial ratios, and are, on average, somewhat flatter than what
was deduced by \citet{bin95}: our median value (see the
3-D illustrations in Fig.~\ref{fig:tree}) is slightly flatter than E3
for the prolate case, and slightly flatter than E4 for the oblate
case.

Can we decide whether the true shapes of our galaxies are more likely
to be oblate or to be prolate? For this purpose, we make use of the
surface brightness test \citep{mar79,ric79}, again following
\citet{bin95}. If dEs were intrinsically oblate spheroids, galaxies that
appear round would be seen face-on and should thus have a lower mean
surface brightness than galaxies that appear flat; the latter would be
seen edge-on. For the prolate case, the inverse relation should be
observed. However, before we can perform this test, we need to take
into account the strong correlation of dE surface brightness with
magnitude \citep[e.g.,][]{bin91}: if, by chance, the few apparently round
galaxies in one of our smaller subsamples would happen to be fainter on
average than the apparently flat ones, this could introduce an
artificial relation of axial ratio with surface brightness. Therefore,
instead of directly using surface brightness like earlier studies did, we
use the surface brightness \emph{offset} from the mean relation of
surface brightness and magnitude. We plot these values, measured in the
$r$ band within $a = 2 a_{{\rm hl},r}$, against axial ratio (measured at
the same semimajor axis, see Sect.~\ref{sec:sub_morpho}) for each dE subsample,
shown in the rightmost column of Fig.~\ref{fig:multiI}. For all
subsamples, a positive correlation of surface brightness offset with
axial ratio can be seen, favoring the oblate model in agreement with
earlier studies \citep[e.g.,][]{mar79,ric79,bin95}. For the ``rounder''
subsamples (top three rows), some additional contribution by prolate
objects might be ``hidden'' within the rather large scatter of surface
brightness offsets at larger axial ratios. We denote these results in
Fig.~\ref{fig:multiI} by the arrows pointing from the
surface brightness test diagram towards the favored intrinsic galaxy
shapes. The arrow size represents the implied contribution from
intrinsically prolate and oblate objects.
 Among the ``flatter'' subsamples (lower three rows), for which
the oblate case is favored, the dE(di)s have the lowest axial ratios,
with a median value of 0.33 (bright dE(nN)s: 0.42, dE(bc)s: 0.44).
%which naively appears to be consistent with the fact that they display disk
%features (Paper I).
The galaxies in these subsamples are thus most likely shaped like thick
disks.

The above considerations needed to be restricted to purely
oblate and purely prolate shapes. However, for all subsamples, a small
part of the deduced (and favored) intrinsic oblate distribution
becomes negative at large axial ratios, trying to account for the low
number of apparently round objects. This implies that most of the
galaxies might actually have triaxial shapes, in accordance with the
conclusions of \citet{bin95}.

\subsection{Subclass distribution within the cluster}
\label{sec:sub_cluster}

While it has been known for some time that nucleated and
non-nucleated dEs have different clustering properties
\citep[e.g.,][]{vdB86, fer89}, this statement has been challenged by
\citet{acsvcs8}, who conjectured that it might just be the result of
a selection bias in the VCC. It therefore appears worth to perform a quantitative
comparison of the distributions of our dE subsamples within the
cluster, and to then proceed with testing the issues raised by
\citet{acsvcs8} in detail. 

\begin{figure}
  \epsscale{0.5}
  \plotone{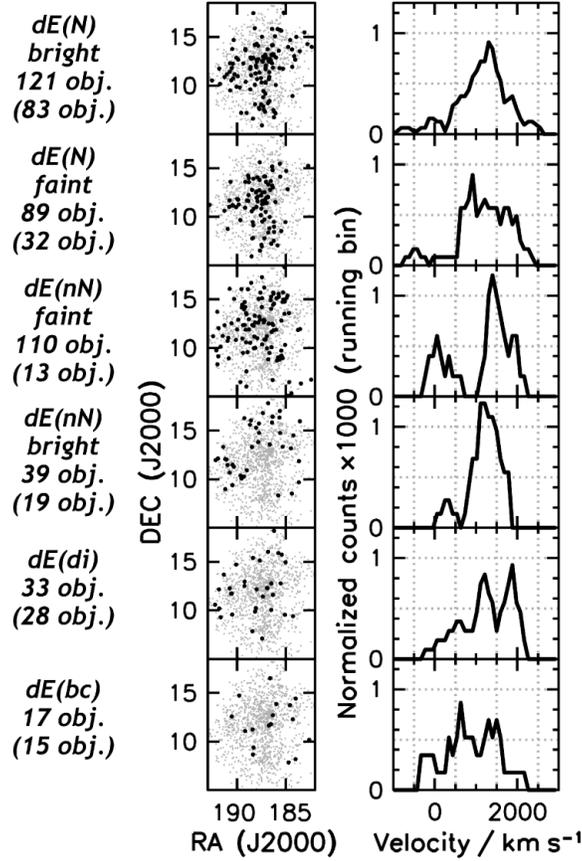}
  \caption{Spatial and velocity distribution.
      In the left column, the number of objects in each of our dE
      subsamples are given. Numbers in parentheses apply to those dEs for
      which heliocentric velocities are available. The middle column
      shows the projected distribution of the dE subsamples (black
      dots) within the cluster. All Virgo cluster member galaxies are
      shown as small grey dots. The right column shows the velocity
      distributions of the dE subsamples. The data are shown as running
      histogram with a bin width of $384\,$km/s, corresponding to the
      semi-interquartile range of the total 193 velocities. Each curve
      is sampled in steps of $96\,$km/s (one quarter of the bin width) and
      normalized to an area of 1.
  }
  \label{fig:multiII}
\end{figure}

The projected spatial distributions of our subsamples are
shown in the middle column of Fig~\ref{fig:multiII}. While both bright
and faint dE(N)s exhibit a rather strong central clustering, the faint
dE(nN)s appear to be only moderately clustered, and the dE(di)s and
dE(bc)s show basically no central clustering. The bright dE(nN)s even seem
to be preferentially located in the outskirts of the cluster.

\begin{figure}
  \epsscale{0.5}
  \plotone{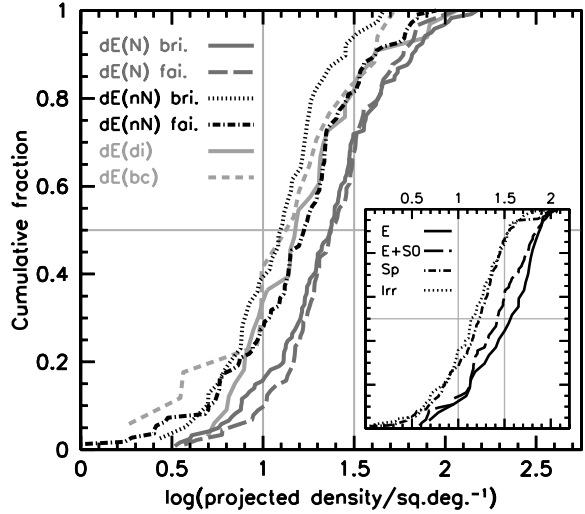}
  \caption{Morphology vs.\ density.
      Cumulative distribution of local projected densities of our
      dE subsamples, and of Hubble types (inset).
      Following
      \citet{dre80} and \citet{bin87}, we define a circular area around each
      galaxy that 
      includes its ten nearest neighbours (independent of galaxy
      type), yielding a projected density (number of galaxies
      per square degree).
  }
  \label{fig:dens}
\end{figure}

To put the above on a more quantitative basis, we present in
Fig.~\ref{fig:dens} the cumulative distribution of each of our
subsamples with respect to local projected density.
Following \citet{dre80} and \citet{bin87}, we define the latter for
each galaxy as the number of objects per 
square degree within a circle that includes the ten nearest neighbours,
independent of galaxy type. Only certain cluster members are
considered. For comparison, we also show the same distributions for
different Hubble types (Fig.~\ref{fig:dens}, inset), i.e., for the
rather strongly centrally clustered giant early-type galaxies, as well
as for the weakly clustered and probably infalling spiral and irregular
galaxies \citep[e.g.,][]{bin87}.

As a confirmation of the impression from the spatial distribution, the
bright dE(nN)s are preferentially found in regions of moderate to lower
density, similar to (and at even slightly lower densities than) the
distribution of irregular galaxies, in accordance with the findings of
\citet{fer89}. This implies that they, as a 
population, are far from being virialized. 
The densities then increase
slightly going from the bright dE(nN)s to the dE(bc)s, dE(di)s, and the
faint dE(nN)s, in this order. Still, all of these are distributed
similarly to the irregular and spiral galaxies in the cluster, again
implying that they are unrelaxed or at least largely unrelaxed galaxy
populations, and confirming the impression from their projected spatial
distribution. In contrast, both bright and faint dE(N)s are located at
larger densities, and display a distribution comparable to the E and S0
galaxies, in agreement with the results of \citet{fer89}. This would
suggest that they are a largely relaxed or at 
least partially relaxed population. Note, however, that the Es
\emph{alone} (without the S0s)
are located at still higher densities. \citet{conI} pointed out that
only the Es appear to be a relaxed galaxy population, while all others,
including the S0s, are not --- thus, the dE(N)s presumably are not
fully relaxed either.

\begin{figure}
  \epsscale{0.5}
  \plotone{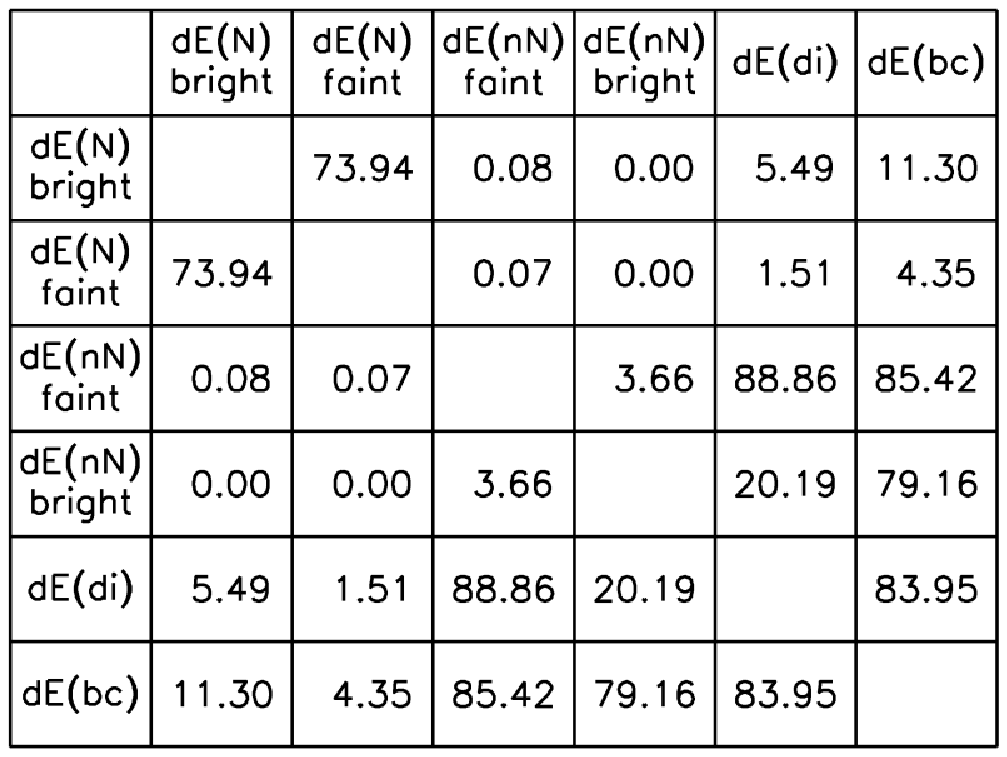}
  \caption{
K-S test results for the comparison of the distributions of our dE
subsamples with respect to local projected density. For each pair of distributions
we give the probability in percent for the null hypothesis that the two
distributions stem from the same underlying distribution function.
  }
  \label{fig:dens_ks}
\end{figure}

We performed statistical pairwise comparisons of the distributions of our dE
subsamples with respect to density, similar as
for the axial ratios in Sect.~\ref{sec:sub_shape}. The K-S test
probabilities for the null hypothesis that two observed distributions
stem from the same underlying distribution are given as percentages in
Fig.~\ref{fig:dens_ks}. Even though the faint dE(nN)s are, among the
``lower-density'' subsamples, closest to the bright and faint dE(N)s, their probability
for having the same underlying distribution is 0.08 and 0.07\%,
respectively. These probabilities are higher for the dE(di)s and
dE(bc)s: although they are located at even lower densities, their
rather small sample sizes let the probability increase as compared to
that of the faint dE(nN)s. Finally, the bright dE(nN)s are located at such low
densities that their K-S test comparison with the dE(N)s yields a
probability of 0.00\%, and that even the comparison with the faint
dE(nN)s only yields a probability of 3.7\% for them having the same
true distribution. Given the morphological differences between
the subsamples, as deduced in Sect.~\ref{sec:sub_shape},
Fig.~\ref{fig:dens} basically shows
a morphology-density relation \emph{within} the dE class.

This view appears to be
corroborated by the distributions of heliocentric velocities
(right column of Fig.~\ref{fig:multiII}) of the dE subsamples: that of
the bright dE(N)s has a single peak and
is fairly symmetric, while especially the faint dE(nN)s, dE(di)s, and
dE(bc)s display rather asymmetric distributions with multiple
peaks. The latter could be interpreted as being a signature of
infalling populations (\citealt{tul84}; \citealt{conI}). However, the
differences between these
velocity distributions are not or only marginally significant
--- the ``most different'' pair of distributions according to the 
K-S test are the bright dE(nN)s and the dE(bc)s, which have a
probability of 6.6\% for the null hypothesis. The main issue here are
the small sample sizes: only a fraction of the galaxies of each
subsample has measured velocities (numbers are given in parentheses in
the left column of Fig.~\ref{fig:multiII}), which are available from
the NED for 193 of our 413 dEs, and, e.g., for only 19 of our 39 bright
dE(nN)s. Similarly, measurements of 
the skew or kurtosis of the distributions do not yield values that
differ significantly from zero.
We can thus only state that
the rather asymmetric, multi-peaked distributions of the faint dE(nN)s,
the dE(di)s, and the dE(bc)s would be consistent with our above
conclusion that they are mostly unrelaxed populations, but that more
velocity data is needed to perform a reliable quantitative comparison
of velocity distributions.

\subsection{Remarks on possible selection biases}
\label{sec:sub_bias}

The different spatial distribution of dE(N)s and dE(nN)s was
long considered a fundamental and well-founded observation, but has
recently been questioned by \citet{acsvcs8}.
These authors argued that
galaxies with high central surface brightness (HSB,
with $\mu_{\rm g,central}\lesssim 20$ mag arcsec$^{-2}$ or $B\lesssim 14.55$) would
have been preferentially classified as non-nucleated in the VCC, which 
 may have lead to a selection bias in
the VCC that artificially relates spatial distribution to nucleus
presence. We test this conjecture by considering the following points:

(1) If the dE(nN)s
      were objects in which nuclei have preferentially gone undetected due to a too
      large central surface brightness, the dE(nN)s' surface
      brightnesses should, on average, be significantly higher
      than those of the dE(N)s. However,
      the mean surface brightness in $r$ within the
      half-light aperture has very similar median values for the bright
      dE(nN)s ($\mu_{r} = 22.65$ mag arcsec$^{-2}$) and the bright
      dE(N)s ($\mu_{r} = 22.63$ mag arcsec$^{-2}$), which makes such a
      bias unlikely. Furthermore, the
      distributions of surface brightnesses of the two subsamples are
      similar --- a K-S test yields a probability of 84\% for the
      null hypothesis that they stem from the same underlying
      distribution. Certainly, measurements of the very central
      surface brightness, which are possible only with high-resolution
      observations, would provide a more direct argument here. However,
      since both nucleated and non-nucleated dEs within a given
      magnitude range have similar surface brightness profiles
      \citep{bin91}, their effective surface brightness and
      central surface brightness are closely correlated.

(2) Only one single galaxy among our 39 bright dE(nN)s (2.5\%) is bright enough to fall among
\citeauthor{acsvcs8}'s definition of a HSB dE. In contrast, 14 of our
      121 bright dE(N)s (12\%) would qualify as HSB dE. Therefore, it
      appears highly unlikely that a significant number of dE(nN)s
      possess nuclei \emph {with similar relative brightnesses as
        those of the dE(N)s} that were not detected by \citet{vcc}.

(3) None of the dEs in \citeauthor{acsvcs8}'s own sample that were previously
classified as non-nucleated, but have now been found to host a weak
nucleus, actually are HSB dEs.

\begin{figure}
  \epsscale{0.3}
  \plotone{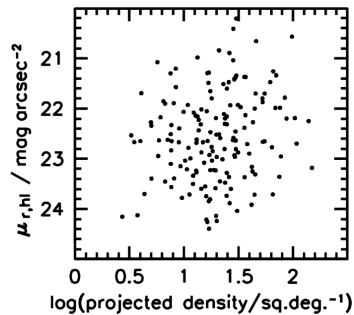}
  \caption{Surface brightness vs.\ density.
      The mean surface brightness in $r$ within the
      half-light aperture is compared to local projected density
      for the combined sample of bright dE(N)s and dE(nN)s, in order to test for
      a possible classification bias as conjectured by \citet{acsvcs8}
      (see text). No correlation is seen, ruling out such a bias.
  }
  \label{fig:densmu}
\end{figure}

(4) Since we are interested in the distributions of our subsamples with
    respect to density in the cluster, we translate \citeauthor{acsvcs8}'s
    conjecture about the \emph{spatial} distribution of the dEs
    into one about the distribution with respect to \emph{density}: if the
    different density distributions of bright dE(N)s and dE(nN)s (see
    above) would primarily be caused by a surface brightness selection
    effect, a significantly larger fraction of the high surface
    brightness objects should be located at lower densities as compared
    to the
    lower surface brightness objects. To test for this possible bias, 
    we plot the
    mean $r$ band surface brightness within the half-light aperture against
    local projected density for the combined sample of bright dE(N)s
    and dE(nN)s (Fig.~\ref{fig:densmu}). No correlation is seen,
    ruling out that such a bias is present in our data.

We point out that it might of course still be the case that most of the
      dE(nN)s host \emph{weak} nuclei that are below the VCC detection
      limit, as discussed in Sect.~\ref{sec:sub_nuc}. However, what is
      at stake here is the question whether a significant number of dE(nN)s
      should already have been classified as dE(N)s \emph{by the VCC},
      and whether this could account for the population differences
      that we find. The above arguments clearly rule out such a
      bias. We can thus conclude that the bright dE(N)s 
      and dE(nN)s are indeed distinct dE subpopulations that differ in
      their clustering properties (Figs.~\ref{fig:dens} and
      \ref{fig:dens_ks}), as well as in their shapes
      (Figs.~\ref{fig:multiI} and \ref{fig:ax_ks}).

%________________________________________________________________

\section{Color analysis}
\label{sec:col}

Since our morphological subdivision of the dEs into several
subpopulations is now established, the next step would obviously be to
compare their stellar population properties. Given that the SDSS imaged
every galaxy in five bands, it should be able to provide some insight
into their stellar content, even though it is basically impossible to
disentangle ages and metallicities with optical broadband photometry
alone. However, the issue is complicated by the fact that the $u$ and
$z$ band images, which would be very important for an analysis of the
stellar content, have a very low S/N (see
Sect.~\ref{sec:data}). It is therefore important to perform a thorough
study of the dE colors and color gradients that properly takes into
account measurement errors and the different S/N levels for objects of
different magnitudes and surface brightnesses. Such a study is beyond
the scope of the present Paper and will be presented in a forthcoming
Paper of this series.

\begin{figure}
  \epsscale{0.5}
  \plotone{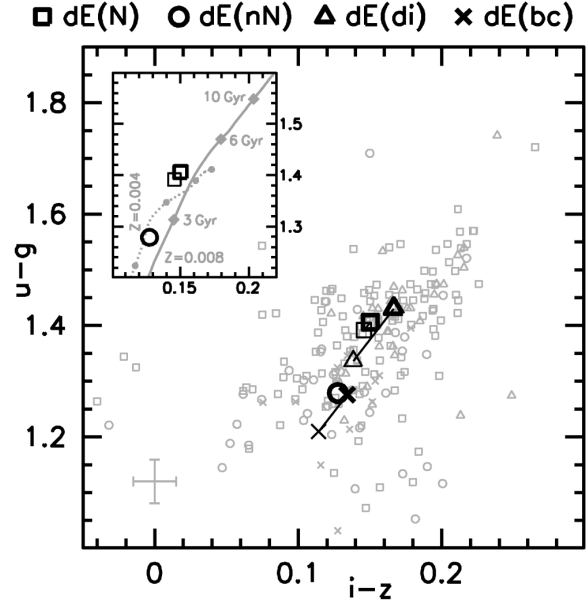}
  \caption{Distribution in color space. Shown are the inner $u-g$ versus $i-z$
  colors, measured within $a = 0.5 a_{{\rm hl},r}$, for
  all dEs brighter than the median $r$ brightness of our full sample,
  $\mr=15.67$ mag, divided into the respective subsamples. Individual
  measurements are shown with small grey symbols (see the legend above
  the diagram). The median value of each subsample is shown as black
  symbol using thick lines. Black symbols drawn with thin lines
  represent the median values corrected for the effect of the
  color-magnitude-relation (see text); lines connect them to the corresponding
  uncorrected values. The correction is chosen to be zero for the
  dE(nN)s. The inset shows again the median values (corrected and
  uncorrected) of the dE(N)s and dE(nN)s, along with two model tracks
  from stellar population synthesis calculations \citep{bc03}. Both
  tracks represent stellar populations formed through a single burst
  of star formation that exponentially decays with time ($\tau =
  0.5$ Gyr), using Padova 2000 isochrones and a Chabrier IMF. The
  grey solid line is for a metallicity $Z=0.008$; age steps are marked
  by the grey diamonds at 3 Gyr, 6 Gyr, and 10 Gyr. The grey dotted
  line is for $Z=0.004$, with age steps marked by the grey circles at 3
  Gyr, 6 Gyr, 10 Gyr, and 14 Gyr; the latter is also the end of the track.
  }
  \label{fig:col}
\end{figure}

Nevertheless, in order to tackle the question about whether the
dE subsamples differ in their color properties, we present in
Fig.~\ref{fig:col} the inner $u-g$ (``age-sensitive'') versus $i-z$
(``metallicity-sensitive'') colors for the bright
($\mr\le15.67$) dEs, measured within an aperture of $a=0.5a_{{\rm
    hl},r}$. This approach guarantees relatively small errors
(typical values are shown in the lower left corner of the Figure) that
need not be taken into account individually. For each dE subsample, we
indicate its median color values with the black symbols drawn with
thick lines.

However, a direct comparison of these values would be
biased by the existence of a color-magnitude relation: if different
subsamples had, on average, significantly different magnitudes, they
would be offset in our color-color diagram even if they followed
exactly the same color-magnitude relation. We therefore compute an
approximate correction for this effect: first, we perform a linear least
squares fit to the color-magnitude relations ($r$ versus $u-g$ and $r$
versus $i-z$) of our full dE sample, clipping one time at 3$\sigma$ and excluding
the dE(bc)s because of their blue inner colors. We then derive the
median $r$ magnitude of each subsample, and use the linear fit to
compute its expected color offset from the sample of dE(nN)s, which we
choose as reference. The so obtained corrected median colors are shown
in Fig.~\ref{fig:col} as black symbols drawn with thin lines, and are
connected with lines to their uncorrected values.

The dE(bc)s exhibit, as expected, the bluest colors of all subsamples,
basically by definition, since we focus here on the inner galaxy
colors. While the corrected colors of the dE(di)s are similar to those
of the dE(nN)s, the dE(N)s are, on average, redder in $i-z$ and
significantly redder in $u-z$. Given the very small color correction
and large sample size of the dE(N)s, this can be considered a robust
result. In the inset shown in Fig.~\ref{fig:col}, we compare the median
values of the dE(nN)s and dE(N)s to two model tracks from stellar
population synthesis calculations \citep{bc03}.
 Both tracks represent stellar populations formed through a single burst
  of star formation that exponentially decays with time ($\tau =
  0.5$ Gyr), using Padova 2000 isochrones and a Chabrier IMF.
 The tracks are curves of constant metallicity (grey solid line:
 $Z=0.008$, grey dotted line $Z=0.004$); ages increase from bottom to
 top and are marked at 3, 6, 10, and 14 Gyr (the latter mark is outside
 of the plot area for the $Z=0.008$ track).
Our measured values
 lie along of the $Z=0.008$ track, illustrating that,
 within the framework of our simplified stellar population models, the
 color difference between dE(nN)s and dE(N)s could be interpreted as a
 difference in age. 
 According to this simple approach, the dE(N)s would
 be, on average, a few Gyr older than the dE(nN)s.
However, the
 measurements also fall roughly along a virtual line connecting the 6
 Gyr points of each model, showing that they might
 also be interpreted as a difference in metallicity.
 While this color offset between the dE(N)s and dE(nN)s would be
 qualitatively consistent with the study by \citet{rak04}, who find the
 dE(N)s in the Coma and Fornax clusters to have older stellar
 populations than the dE(nN)s, reliable conclusions need to await a
 more comprehensive color study of our dEs.

%________________________________________________________________

\section{Discussion}
\label{sec:discussion}

\subsection{Interrelations between subclasses}
\label{sec:sub_inter}

The bright dE(nN)s and dE(di)s are both
unrelaxed populations of relatively bright dEs shaped like thick
disks. This also applies to the dE(bc)s, which could thus be candidates
for being the direct progenitors of the former: the presently blue
centers of the dE(bc)s will evolve to typical dE colors within 1 Gyr or less
after the cessation of star formation (Paper II). Therefore, the bright
dE(nN)s and dE(di)s could constitute those disk-shaped dEs where
central star formation has already ceased.
To test this hypothesis, we make the
following considerations. There are 39 bright dE(nN)s, as well as
30 dE(di)s with $\mr \le 15.67$ mag, 7 of which are non-nucleated. This adds
up to 69 ``non-star-forming, disk-shaped dEs'',  23 (33\%) of which are
nucleated. Among the dE(bc)s there are 15 galaxies with $\mr \le 15.67$
mag, 6 (40\%) of which are nucleated. Thus, the fraction of nucleated
galaxies would be compatible with our hypothesis within the errors,
with the caveat that nuclei might still form in the centers of some
dE(bc)s (see \citealt{oh00} and Paper II), which would raise the
nucleated fraction of the dE(bc)s.

Now, 43\% of the non-star-forming, disk-shaped dEs are dE(di)s, i.e., show disk
\emph{features} (not only an overall disk \emph{shape}). If
the dE(bc)s would contain the same fraction of galaxies that display
disk features, we would expect 6.5 such objects among the 15
dE(bc)s, with a standard deviation of 1.9. The observed number of 4
lies within 1.3\,$\sigma$ of the expected value and could thus still be
reconciled with the above picture.
However, since not only the dE(di)s, but also the bright dE(nN)s
are disk-shaped, why do the latter not display disk \emph{features}
like the dE(di)s? This could either indicate
a correlation between the
presence of a significantly bright nucleus and the presence of disk
substructure, or it could imply that there is more than one
formation path towards disk-shaped dEs.

%________________________________________________________________

\subsection{Formation mechanisms}
\label{sec:sub_formation}

If dEs originated from galaxies that fell into the cluster,
how long ago could this infall have taken place?
\citet{conI} derived a two-body relaxation time for the Virgo dEs of
much more than a Hubble time.
Even violent relaxation, which could apply for the case of infalling or
merging groups, would take at least a few crossing times $t_{\rm cr}$,
with $t_{\rm cr}\approx 1.7$ Gyr for the Virgo cluster \citep{bos06}.
Therefore, the majority of dE(N)s or their progenitors should have experienced
infall in the earliest phases of the Virgo cluster (which is a rather
young structure, see \citealt{bin87} and \citealt{arn04}), or they could have
formed in dark matter halos along with the cluster itself.
All other dE subclasses are largely unrelaxed populations,
implying that they have formed later than the dE(N)s, probably from
(continuous) infall of progenitor galaxies. Our color analysis in
Sect.~\ref{sec:col} would support this view, since it finds that the
inner colors of the dE(N)s can be interpreted with an older stellar
population than the dE(nN)s. This would be expected if one assumes
that the progenitor galaxies had been forming stars until their infall
into the cluster, resulting in a younger stellar population on average
in the case of a later infall (neglecting possible metallicity
differences). However, as stressed in Sect.~\ref{sec:col}, robust
conclusions need to await a more detailed multicolor study of our dEs.

The galaxy harassment scenario \citep{moo96} describes the structural
transformation of a late-type spiral into a spheroidal system through
strong tidal interactions with massive cluster galaxies.  A thick
stellar disk may survive 
and form a
bar and spiral features that can be retained for some time, depending
on the tidal heating of the galaxy \citep{mas05}. Harassment could thus
form disk-shaped dEs, and dE(di)s in particular. Moreover, it predicts
gas to be funneled to the center and form a density excess there
\citep{moo98}, which would be well suited to explain the central star
formation in the dE(bc)s.
Therefore, it appears possible that harassment could form disk-shaped
dEs that first appear as dE(bc)s and then passively evolve into
dE(di)s and bright dE(nN)s as their star formation ceases (see
Sect.~\ref{sec:sub_inter}). It
might also provide a way to form the fainter non-nucleated dEs,
assuming that the tidal forces have a stronger effect on the shape of
less massive galaxies, resulting in rounder objects on
average. However, in order to explain all these
subclasses by a single process, one would need to invoke 
a correlation between the presence of a nucleus and of
disk features, as discussed in Sect.~\ref{sec:sub_inter}.

Ram-pressure
stripping \citep{gun72} of dwarf irregulars (dIrrs) could be
responsible for the fact that the
disk-shaped bright dE(nN)s do not show disk \emph{features}
like the dE(di)s:
dIrrs typically have no nucleus, and 
ram-pressure stripping exerts much less perturbing forces than a
violent process like harassment, thus probably not triggering the
formation of bars or spiral arms.
Commonly discussed problems with this scenario are the metallicity
offset between dEs and dIrrs \citep{thu85,ric98,gre03} and the too
strong fading of dIrrs after cessation of star formation
\citep{bot86}. Also, the flattening
distribution of Virgo cluster dIrrs -- with intrinsic (primary) axial
ratios $\ge$ 0.5 for most galaxies \citep{bin95} -- is not quite
like that of our bright dE(nN)s.
On the other hand, significant mass loss due to stripped gas 
might affect
the stellar configuration of the
galaxies and could thus possibly account for the difference.
Moreover, the flattening distribution of the dIrrs is similar to that of
  the \emph{faint} dE(nN)s (cf.\ Fig.~9 of \citealt{bin95}), suggesting that
  these -- and possibly not the \emph{bright} dE(nN)s -- might be stripped
  dIrrs.

Tidally induced star formation of dIrrs might be able to overcome the
problems of the ram-pressure stripping scenario: the initially lower
metallicity and surface brightness of a dIrr are increased by several bursts
of star formation \citep{dav88}, during which the galaxy appears as 
blue compact dwarf (BCD). After the last BCD phase it fades to become a dE,
thereby providing an explanation how BCDs could be dE progenitors, which has
frequently been discussed (e.g., \citealt{bot86}; \citealt{pap96}; \citealt{gre97};
Paper II). The last star
formation burst might occur in the central region, 
consistent with the appearance of the dE(bc)s. 

In addition to the number of possible formation scenarios, the role of
the nuclei provides another unknown element.
If dE(N)s and dE(nN)s would actually form a continuum of dEs
with respect to relative nucleus brightness
as suggested by \citet{gra05},
their significantly different population properties could be interpreted with a
correlation between
relative nucleus brightness
and host galaxy evolution.
Such a correlation could, for example, be provided by
nucleus formation through coalescence of globular clusters (GCs): the infall
and merging of \emph{several} GCs -- resulting in a
rather bright nucleus like in a dE(N) -- takes many Gyr \citep{oh00}, consistent with
the dE(N)s being in place since long. The dE(nN)s, on the other hand, were
probably formed more recently, leaving time for only one or two GCs,
or none at all, to sink to the center.

\subsection{Remarks on previous work}
\label{sec:sub_compare}

Results similar to ours were derived by \citet{fer89},
who also subdivided 
Virgo and Fornax cluster dEs with respect to magnitude and the presence
or absence of a nucleus. In accordance with our results, they found
that the dE(N)s are centrally clustered like E and S0 galaxies, while
the bright dE(nN)s are distributed like spiral and irregular
galaxies. They also found the axial ratios of the bright dE(nN)s to be
flatter than those of the dE(N)s.

However, despite these similar
results, their magnitude selection of ``bright'' and ``faint''
subsamples is actually quite 
different from ours. We initially selected only dEs with $\mb\le 18.0$
mag (the completeness limit of the VCC), yielding a sample range of
about 4.5 mag in $B$, and then subdivided our full sample at its median $r$
magnitude. In contrast to that, \citet{fer89} included VCC galaxies with
$\mb< 17.5$ in their \emph{bright} subsample, which therefore still
spans a range of 4 mag. Their \emph{faint} subsample contains VCC galaxies
with $\mb> 18.4$, which are not included in our study and already lie
within the luminosity regime of Local Group
dwarf spheroidals \citep[e.g.,][]{gre03}. Therefore, their and our
study can be considered complementary to some
extent, in the sense that we probe different luminosity regimes with
our respective subsample definitions.

%________________________________________________________________

\section{Conclusions}
\label{sec:summary}

We have presented a quantitative analysis of the intrinsic shapes and
spatial distributions of various subsamples of
Virgo cluster early-type dwarfs (dEs): bright and
faint (non-)nucleated dEs (dE(N)s and dE(nN)s), dEs with disk features
(dE(di)s), and dEs with blue centers dE(bc)s). The dE(bc)s, dE(di)s,
and bright dE(nN)s are shaped like thick disks and show basically no
central clustering, indicating that they are unrelaxed populations
that probably formed from infalling progenitor galaxies. As opposed to
that, the dE(N)s (both bright and faint) are a fairly relaxed
population of spheroidal galaxies, though an oblate intrinsic shape is
favored for them as well. The faint dE(nN)s appear to be somewhat
intermediate: their shapes are similar to the dE(N)s, but they form a
largely unrelaxed population as derived from their clustering
properties. 
 Taken together, these results define a
morphology-density relation \emph{within} the dE class.

Given that \citet{fer89} derived similar results for both Virgo and
Fornax cluster galaxies, it is also clear that this zoo of different dE
subclasses is not only specific to the Virgo cluster. Similarly, a
significant number of Coma cluster dEs show a two-component profile
and are flatter than the normal dEs \citep{agu05}. Moreover,
\citet{rak04} found the dE(N)s in Coma and Fornax to have older stellar
populations than the dE(nN)s, consistent with our color analysis of the
Virgo cluster dEs. Thus, although the relative proportions of the dE
subclasses might vary between the dynamically different Virgo, Coma,
and Fornax clusters, the dE variety itself is probably similar in any
galaxy cluster in the present epoch. We thus consider it important that
future studies of dEs do not intermingle the different subclasses, but
instead compare their properties with each other, e.g., their stellar content or kinematical
structure. This will eventually lead to pinning down the actual
significance of the various suggested formation paths, thereby
unveiling an important part of galaxy cluster formation and evolution.

%________________________________________________________________

\acknowledgements
    We thank the referees for useful suggestions that
    helped us improve the paper. We gratefully acknowledge support by
    the Swiss National Science Foundation through grants number
    200020-105260 and 200020-105535. We thank J.\ Gallagher for
    stimulating discussions, and W.\ L\"offler for his untiring computer
    support.

    This study would not have been possible without the wealth of publicly
    available data from the SDSS.    
    Funding for the SDSS has been provided by the Alfred
    P. Sloan Foundation, the Participating Institutions, the National
    Science Foundation, the U.S. Department of Energy, the National
    Aeronautics and Space Administration, the Japanese Monbukagakusho,
    the Max Planck Society, and the Higher Education Funding Council
    for England. The SDSS Web Site is http://www.sdss.org/. 

    The SDSS is managed by the Astrophysical Research Consortium for
    the Participating Institutions. The Participating Institutions are
    the American Museum of Natural History, Astrophysical Institute
    Potsdam, University of Basel, Cambridge University, Case Western
    Reserve University, University of Chicago, Drexel University,
    Fermilab, the Institute for Advanced Study, the Japan Participation
    Group, Johns Hopkins University, the Joint Institute for Nuclear
    Astrophysics, the Kavli Institute for Particle Astrophysics and
    Cosmology, the Korean Scientist Group, the Chinese Academy of
    Sciences (LAMOST), Los Alamos National Laboratory, the
    Max-Planck-Institute for Astronomy (MPIA), the Max-Planck-Institute
    for Astrophysics (MPA), New Mexico State University, Ohio State
    University, University of Pittsburgh, University of Portsmouth,
    Princeton University, the United States Naval Observatory, and the
    University of Washington. 

    This research has made use of NASA's Astrophysics Data
    System Bibliographic Services, and of the NASA/IPAC Extragalactic
    Database (NED) which is operated by the Jet Propulsion Laboratory,
    California Institute of Technology, under contract with the
    National Aeronautics and Space Administration.

%________________________________________________________________
 
%\clearpage

%\bibliography{virgodE}
\bibliographystyle{apj}

%________________________________________________________________

\clearpage

\begin{deluxetable}{lll|lll|lll}
%\tabletypesize{\scriptsize}
  \tablecaption{Subclass assignment. \label{tab:zoo}}
  \tablehead{
    & & \colhead{$\mr$} & & & \colhead{$\mr$} & & & \colhead{$\mr$} \\
    \colhead{VCC} & \colhead{Subclass} & (mag)
    & \colhead{VCC} & \colhead{Subclass} & (mag)
    & \colhead{VCC} & \colhead{Subclass} & (mag)
  }
  \startdata
0009 & dE(N)$_{\rm bright}$ &  12.94 &  1222 & dE(N)$_{\rm bright}$ &  15.20   &  0747 & dE(N)$_{\rm faint}$ &  16.17    \\ 
0033 & dE(N)$_{\rm bright}$ &  14.26 &  1238 & dE(N)$_{\rm bright}$ &  14.49   &  0755 & dE(N)$_{\rm faint}$ &  15.77    \\ 
0050 & dE(N)$_{\rm bright}$ &  15.12 &  1254 & dE(N)$_{\rm bright}$ &  13.92   &  0756 & dE(N)$_{\rm faint}$ &  16.31    \\ 
0109 & dE(N)$_{\rm bright}$ &  15.29 &  1261 & dE(N)$_{\rm bright}$ &  12.62   &  0779 & dE(N)$_{\rm faint}$ &  16.89    \\ 
0158 & dE(N)$_{\rm bright}$ &  14.87 &  1308 & dE(N)$_{\rm bright}$ &  14.59   &  0795 & dE(N)$_{\rm faint}$ &  16.52    \\ 
0178 & dE(N)$_{\rm bright}$ &  14.61 &  1311 & dE(N)$_{\rm bright}$ &  15.09   &  0810 & dE(N)$_{\rm faint}$ &  15.79    \\ 
0200 & dE(N)$_{\rm bright}$ &  14.02 &  1333 & dE(N)$_{\rm bright}$ &  15.65   &  0812 & dE(N)$_{\rm faint}$ &  15.89    \\ 
0227 & dE(N)$_{\rm bright}$ &  14.12 &  1353 & dE(N)$_{\rm bright}$ &  15.58   &  0855 & dE(N)$_{\rm faint}$ &  16.44    \\ 
0230 & dE(N)$_{\rm bright}$ &  14.72 &  1355 & dE(N)$_{\rm bright}$ &  13.50   &  0877 & dE(N)$_{\rm faint}$ &  16.86    \\ 
0235 & dE(N)$_{\rm bright}$ &  15.56 &  1384 & dE(N)$_{\rm bright}$ &  15.38   &  0896 & dE(N)$_{\rm faint}$ &  16.66    \\ 
0273 & dE(N)$_{\rm bright}$ &  15.41 &  1386 & dE(N)$_{\rm bright}$ &  13.80   &  0920 & dE(N)$_{\rm faint}$ &  16.20    \\ 
0319 & dE(N)$_{\rm bright}$ &  14.31 &  1389 & dE(N)$_{\rm bright}$ &  15.11   &  0933 & dE(N)$_{\rm faint}$ &  15.76    \\ 
0437 & dE(N)$_{\rm bright}$ &  13.13 &  1400 & dE(N)$_{\rm bright}$ &  14.86   &  0972 & dE(N)$_{\rm faint}$ &  15.78    \\ 
0452 & dE(N)$_{\rm bright}$ &  15.02 &  1407 & dE(N)$_{\rm bright}$ &  14.14   &  0974 & dE(N)$_{\rm faint}$ &  15.69    \\ 
0510 & dE(N)$_{\rm bright}$ &  14.12 &  1420 & dE(N)$_{\rm bright}$ &  15.48   &  0977 & dE(N)$_{\rm faint}$ &  17.11    \\ 
0545 & dE(N)$_{\rm bright}$ &  14.48 &  1431 & dE(N)$_{\rm bright}$ &  13.37   &  0997 & dE(N)$_{\rm faint}$ &  17.04    \\ 
0560 & dE(N)$_{\rm bright}$ &  15.61 &  1441 & dE(N)$_{\rm bright}$ &  15.38   &  1044 & dE(N)$_{\rm faint}$ &  16.16    \\ 
0592 & dE(N)$_{\rm bright}$ &  15.43 &  1446 & dE(N)$_{\rm bright}$ &  15.03   &  1059 & dE(N)$_{\rm faint}$ &  17.10    \\ 
0634 & dE(N)$_{\rm bright}$ &  12.71 &  1451 & dE(N)$_{\rm bright}$ &  15.26   &  1064 & dE(N)$_{\rm faint}$ &  16.44    \\ 
0684 & dE(N)$_{\rm bright}$ &  15.10 &  1453 & dE(N)$_{\rm bright}$ &  13.25   &  1065 & dE(N)$_{\rm faint}$ &  15.70    \\ 
0695 & dE(N)$_{\rm bright}$ &  15.24 &  1491 & dE(N)$_{\rm bright}$ &  14.11   &  1076 & dE(N)$_{\rm faint}$ &  16.22    \\ 
0711 & dE(N)$_{\rm bright}$ &  15.46 &  1497 & dE(N)$_{\rm bright}$ &  15.03   &  1099 & dE(N)$_{\rm faint}$ &  16.74    \\ 
0725 & dE(N)$_{\rm bright}$ &  14.90 &  1503 & dE(N)$_{\rm bright}$ &  14.38   &  1105 & dE(N)$_{\rm faint}$ &  15.79    \\ 
0745 & dE(N)$_{\rm bright}$ &  13.54 &  1539 & dE(N)$_{\rm bright}$ &  14.88   &  1115 & dE(N)$_{\rm faint}$ &  16.26    \\ 
0750 & dE(N)$_{\rm bright}$ &  14.15 &  1549 & dE(N)$_{\rm bright}$ &  13.85   &  1119 & dE(N)$_{\rm faint}$ &  16.29    \\ 
0753 & dE(N)$_{\rm bright}$ &  15.11 &  1561 & dE(N)$_{\rm bright}$ &  14.97   &  1120 & dE(N)$_{\rm faint}$ &  16.09    \\ 
0762 & dE(N)$_{\rm bright}$ &  14.81 &  1563 & dE(N)$_{\rm bright}$ &  15.45   &  1123 & dE(N)$_{\rm faint}$ &  15.74    \\ 
0765 & dE(N)$_{\rm bright}$ &  15.41 &  1565 & dE(N)$_{\rm bright}$ &  15.62   &  1137 & dE(N)$_{\rm faint}$ &  16.89    \\ 
0786 & dE(N)$_{\rm bright}$ &  14.06 &  1567 & dE(N)$_{\rm bright}$ &  13.80   &  1191 & dE(N)$_{\rm faint}$ &  16.64    \\ 
0790 & dE(N)$_{\rm bright}$ &  14.96 &  1649 & dE(N)$_{\rm bright}$ &  14.75   &  1207 & dE(N)$_{\rm faint}$ &  16.06    \\ 
0808 & dE(N)$_{\rm bright}$ &  15.47 &  1661 & dE(N)$_{\rm bright}$ &  14.91   &  1210 & dE(N)$_{\rm faint}$ &  16.47    \\ 
0815 & dE(N)$_{\rm bright}$ &  14.96 &  1669 & dE(N)$_{\rm bright}$ &  15.44   &  1212 & dE(N)$_{\rm faint}$ &  16.05    \\ 
0816 & dE(N)$_{\rm bright}$ &  13.64 &  1674 & dE(N)$_{\rm bright}$ &  15.10   &  1225 & dE(N)$_{\rm faint}$ &  16.31    \\ 
0823 & dE(N)$_{\rm bright}$ &  14.85 &  1711 & dE(N)$_{\rm bright}$ &  15.35   &  1240 & dE(N)$_{\rm faint}$ &  16.63    \\ 
0824 & dE(N)$_{\rm bright}$ &  15.38 &  1755 & dE(N)$_{\rm bright}$ &  14.86   &  1246 & dE(N)$_{\rm faint}$ &  16.84    \\ 
0846 & dE(N)$_{\rm bright}$ &  15.17 &  1773 & dE(N)$_{\rm bright}$ &  15.30   &  1264 & dE(N)$_{\rm faint}$ &  15.68    \\ 
0871 & dE(N)$_{\rm bright}$ &  14.57 &  1796 & dE(N)$_{\rm bright}$ &  15.64   &  1268 & dE(N)$_{\rm faint}$ &  16.29    \\ 
0916 & dE(N)$_{\rm bright}$ &  14.88 &  1803 & dE(N)$_{\rm bright}$ &  14.82   &  1296 & dE(N)$_{\rm faint}$ &  16.18    \\ 
0928 & dE(N)$_{\rm bright}$ &  15.21 &  1826 & dE(N)$_{\rm bright}$ &  14.79   &  1302 & dE(N)$_{\rm faint}$ &  16.94    \\ 
0929 & dE(N)$_{\rm bright}$ &  12.51 &  1828 & dE(N)$_{\rm bright}$ &  14.27   &  1307 & dE(N)$_{\rm faint}$ &  17.19    \\ 
0931 & dE(N)$_{\rm bright}$ &  15.49 &  1861 & dE(N)$_{\rm bright}$ &  13.22   &  1317 & dE(N)$_{\rm faint}$ &  17.25    \\ 
0936 & dE(N)$_{\rm bright}$ &  14.70 &  1876 & dE(N)$_{\rm bright}$ &  14.23   &  1366 & dE(N)$_{\rm faint}$ &  16.15    \\ 
0940 & dE(N)$_{\rm bright}$ &  13.78 &  1881 & dE(N)$_{\rm bright}$ &  15.25   &  1369 & dE(N)$_{\rm faint}$ &  16.02    \\ 
0949 & dE(N)$_{\rm bright}$ &  14.31 &  1886 & dE(N)$_{\rm bright}$ &  14.52   &  1373 & dE(N)$_{\rm faint}$ &  16.67    \\ 
0965 & dE(N)$_{\rm bright}$ &  14.54 &  1897 & dE(N)$_{\rm bright}$ &  13.49   &  1396 & dE(N)$_{\rm faint}$ &  16.29    \\ 
0992 & dE(N)$_{\rm bright}$ &  15.35 &  1909 & dE(N)$_{\rm bright}$ &  15.47   &  1399 & dE(N)$_{\rm faint}$ &  15.85    \\ 
1005 & dE(N)$_{\rm bright}$ &  15.23 &  1919 & dE(N)$_{\rm bright}$ &  15.64   &  1402 & dE(N)$_{\rm faint}$ &  17.27    \\ 
1069 & dE(N)$_{\rm bright}$ &  15.58 &  1936 & dE(N)$_{\rm bright}$ &  14.90   &  1418 & dE(N)$_{\rm faint}$ &  16.37    \\ 
1073 & dE(N)$_{\rm bright}$ &  13.19 &  1942 & dE(N)$_{\rm bright}$ &  15.31   &  1481 & dE(N)$_{\rm faint}$ &  17.10    \\ 
1075 & dE(N)$_{\rm bright}$ &  14.01 &  1945 & dE(N)$_{\rm bright}$ &  13.98   &  1495 & dE(N)$_{\rm faint}$ &  16.69    \\ 
1079 & dE(N)$_{\rm bright}$ &  15.67 &  1991 & dE(N)$_{\rm bright}$ &  14.52   &  1496 & dE(N)$_{\rm faint}$ &  17.20    \\ 
1087 & dE(N)$_{\rm bright}$ &  12.59 &  2012 & dE(N)$_{\rm bright}$ &  13.75   &  1498 & dE(N)$_{\rm faint}$ &  15.91    \\ 
1092 & dE(N)$_{\rm bright}$ &  15.64 &  2045 & dE(N)$_{\rm bright}$ &  14.95   &  1509 & dE(N)$_{\rm faint}$ &  15.77    \\ 
1093 & dE(N)$_{\rm bright}$ &  15.53 &  2049 & dE(N)$_{\rm bright}$ &  15.47   &  1519 & dE(N)$_{\rm faint}$ &  16.57    \\ 
1101 & dE(N)$_{\rm bright}$ &  15.11 &  2083 & dE(N)$_{\rm bright}$ &  14.74   &  1523 & dE(N)$_{\rm faint}$ &  16.61    \\ 
1104 & dE(N)$_{\rm bright}$ &  14.51 &  0029 & dE(N)$_{\rm faint}$ &  16.58    &  1531 & dE(N)$_{\rm faint}$ &  16.62    \\ 
1107 & dE(N)$_{\rm bright}$ &  14.56 &  0330 & dE(N)$_{\rm faint}$ &  15.82    &  1533 & dE(N)$_{\rm faint}$ &  16.80    \\ 
1122 & dE(N)$_{\rm bright}$ &  13.96 &  0372 & dE(N)$_{\rm faint}$ &  17.48    &  1603 & dE(N)$_{\rm faint}$ &  16.47    \\ 
1151 & dE(N)$_{\rm bright}$ &  15.54 &  0394 & dE(N)$_{\rm faint}$ &  16.70    &  1604 & dE(N)$_{\rm faint}$ &  15.99    \\ 
1164 & dE(N)$_{\rm bright}$ &  15.24 &  0503 & dE(N)$_{\rm faint}$ &  16.43    &  1606 & dE(N)$_{\rm faint}$ &  16.61    \\ 
1167 & dE(N)$_{\rm bright}$ &  14.15 &  0505 & dE(N)$_{\rm faint}$ &  16.66    &  1609 & dE(N)$_{\rm faint}$ &  16.19    \\ 
1172 & dE(N)$_{\rm bright}$ &  15.32 &  0539 & dE(N)$_{\rm faint}$ &  15.85    &  1616 & dE(N)$_{\rm faint}$ &  15.79    \\ 
1173 & dE(N)$_{\rm bright}$ &  15.28 &  0554 & dE(N)$_{\rm faint}$ &  16.12    &  1642 & dE(N)$_{\rm faint}$ &  16.58    \\ 
1185 & dE(N)$_{\rm bright}$ &  14.44 &  0632 & dE(N)$_{\rm faint}$ &  16.53    &  1677 & dE(N)$_{\rm faint}$ &  16.07    \\ 
1213 & dE(N)$_{\rm bright}$ &  15.49 &  0706 & dE(N)$_{\rm faint}$ &  16.69    &  1683 & dE(N)$_{\rm faint}$ &  15.71    \\ 
1218 & dE(N)$_{\rm bright}$ &  15.09 &  0746 & dE(N)$_{\rm faint}$ &  17.30    &  1767 & dE(N)$_{\rm faint}$ &  15.68    
  \enddata
  \tablecomments{
Classification of a dE as nucleated or non-nucleated is provided by the
  VCC \citep{vcc}. Of the dEs that do not display disk substructure or
  a blue center, those with a small uncertainty on the presence of a nucleus (``N:'') were
  included in the nucleated subclass, while those with a larger
  uncertainty (``N?'', ``Npec'') were not assigned to any subclass
  (entry ``---'' as subclass),
  but were excluded from all comparisons of dE subclasses. 
  Objects VCC 0218, 0308, 1684, and 1779 are dE(bc)s with disk features.
  }
\end{deluxetable}
 
\setcounter{table}{0}

\begin{deluxetable}{lll|lll|lll}
%\tabletypesize{\scriptsize}
  \tablecaption{Continued. \label{tab:zoo_part2}}
  \tablehead{
    & & \colhead{$\mr$} & & & \colhead{$\mr$} & & & \colhead{$\mr$} \\
    \colhead{VCC} & \colhead{Subclass} & (mag)
    & \colhead{VCC} & \colhead{Subclass} & (mag)
    & \colhead{VCC} & \colhead{Subclass} & (mag)
  }
  \startdata
  1785 & dE(N)$_{\rm faint}$ &  16.96    & 0561 & dE(nN)$_{\rm faint}$ &  16.73     & 1815 & dE(nN)$_{\rm faint}$ &  16.64          \\
  1794 & dE(N)$_{\rm faint}$ &  16.82    & 0594 & dE(nN)$_{\rm faint}$ &  16.34     & 1843 & dE(nN)$_{\rm faint}$ &  16.64          \\
  1812 & dE(N)$_{\rm faint}$ &  16.64    & 0600 & dE(nN)$_{\rm faint}$ &  18.47     & 1867 & dE(nN)$_{\rm faint}$ &  16.62          \\
  1831 & dE(N)$_{\rm faint}$ &  17.38    & 0622 & dE(nN)$_{\rm faint}$ &  17.65     & 1915 & dE(nN)$_{\rm faint}$ &  16.09          \\
  1879 & dE(N)$_{\rm faint}$ &  16.08    & 0652 & dE(nN)$_{\rm faint}$ &  16.76     & 1950 & dE(nN)$_{\rm faint}$ &  15.67          \\
  1891 & dE(N)$_{\rm faint}$ &  16.16    & 0668 & dE(nN)$_{\rm faint}$ &  16.06     & 1964 & dE(nN)$_{\rm faint}$ &  17.23          \\
  1928 & dE(N)$_{\rm faint}$ &  16.56    & 0687 & dE(nN)$_{\rm faint}$ &  17.31     & 1971 & dE(nN)$_{\rm faint}$ &  15.69          \\
  1951 & dE(N)$_{\rm faint}$ &  15.85    & 0748 & dE(nN)$_{\rm faint}$ &  16.26     & 1983 & dE(nN)$_{\rm faint}$ &  16.06          \\
  1958 & dE(N)$_{\rm faint}$ &  15.99    & 0760 & dE(nN)$_{\rm faint}$ &  16.32     & 2011 & dE(nN)$_{\rm faint}$ &  16.40          \\
  1980 & dE(N)$_{\rm faint}$ &  15.94    & 0761 & dE(nN)$_{\rm faint}$ &  16.33     & 2032 & dE(nN)$_{\rm faint}$ &  16.08          \\
  2014 & dE(N)$_{\rm faint}$ &  16.04    & 0769 & dE(nN)$_{\rm faint}$ &  16.20     & 2043 & dE(nN)$_{\rm faint}$ &  17.07          \\
  2088 & dE(N)$_{\rm faint}$ &  16.24    & 0775 & dE(nN)$_{\rm faint}$ &  17.17     & 2051 & dE(nN)$_{\rm faint}$ &  16.54          \\
0108 & dE(nN)$_{\rm bright}$ &  15.08    & 0777 & dE(nN)$_{\rm faint}$ &  16.61     & 2054 & dE(nN)$_{\rm faint}$ &  15.76          \\
0115 & dE(nN)$_{\rm bright}$ &  15.63    & 0791 & dE(nN)$_{\rm faint}$ &  16.22     & 2061 & dE(nN)$_{\rm faint}$ &  16.68          \\
0118 & dE(nN)$_{\rm bright}$ &  15.41    & 0803 & dE(nN)$_{\rm faint}$ &  17.87     & 2063 & dE(nN)$_{\rm faint}$ &  16.54          \\
0209 & dE(nN)$_{\rm bright}$ &  14.18    & 0839 & dE(nN)$_{\rm faint}$ &  16.80     & 2074 & dE(nN)$_{\rm faint}$ &  16.58          \\
0236 & dE(nN)$_{\rm bright}$ &  15.23    & 0840 & dE(nN)$_{\rm faint}$ &  17.09     & 2081 & dE(nN)$_{\rm faint}$ &  16.17          \\
0261 & dE(nN)$_{\rm bright}$ &  15.31    & 0861 & dE(nN)$_{\rm faint}$ &  16.73     & 0535 & --- &  15.69                           \\
0461 & dE(nN)$_{\rm bright}$ &  15.38    & 0863 & dE(nN)$_{\rm faint}$ &  16.73     & 1348 & --- &  14.15                           \\
0543 & dE(nN)$_{\rm bright}$ &  13.35    & 0878 & dE(nN)$_{\rm faint}$ &  16.04     & 1489 & --- &  15.07                           \\
0551 & dE(nN)$_{\rm bright}$ &  15.17    & 0926 & dE(nN)$_{\rm faint}$ &  16.06     & 1857 & --- &  13.92                           \\
0563 & dE(nN)$_{\rm bright}$ &  14.80    & 0962 & dE(nN)$_{\rm faint}$ &  16.07     & 0216 & dE(di) &  14.31                        \\
0611 & dE(nN)$_{\rm bright}$ &  15.50    & 0976 & dE(nN)$_{\rm faint}$ &  16.93     & 0389 & dE(di) &  13.09                        \\
0794 & dE(nN)$_{\rm bright}$ &  13.88    & 1034 & dE(nN)$_{\rm faint}$ &  17.42     & 0407 & dE(di) &  13.74                        \\
0817 & dE(nN)$_{\rm bright}$ &  13.09    & 1039 & dE(nN)$_{\rm faint}$ &  16.29     & 0490 & dE(di) &  13.00                        \\
0917 & dE(nN)$_{\rm bright}$ &  14.54    & 1089 & dE(nN)$_{\rm faint}$ &  16.98     & 0523 & dE(di) &  12.52                        \\
0982 & dE(nN)$_{\rm bright}$ &  15.26    & 1124 & dE(nN)$_{\rm faint}$ &  16.75     & 0608 & dE(di) &  13.56                        \\
1180 & dE(nN)$_{\rm bright}$ &  15.41    & 1129 & dE(nN)$_{\rm faint}$ &  16.87     & 0751 & dE(di) &  13.74                        \\
1323 & dE(nN)$_{\rm bright}$ &  15.49    & 1132 & dE(nN)$_{\rm faint}$ &  15.85     & 0788 & dE(di) &  15.33                        \\
1334 & dE(nN)$_{\rm bright}$ &  14.69    & 1149 & dE(nN)$_{\rm faint}$ &  16.78     & 0854 & dE(di) &  16.71                        \\
1351 & dE(nN)$_{\rm bright}$ &  14.91    & 1153 & dE(nN)$_{\rm faint}$ &  16.51     & 0856 & dE(di) &  13.38                        \\
1417 & dE(nN)$_{\rm bright}$ &  15.06    & 1209 & dE(nN)$_{\rm faint}$ &  16.55     & 0990 & dE(di) &  13.70                        \\
1528 & dE(nN)$_{\rm bright}$ &  13.67    & 1223 & dE(nN)$_{\rm faint}$ &  15.74     & 1010 & dE(di) &  12.72                        \\
1553 & dE(nN)$_{\rm bright}$ &  15.53    & 1224 & dE(nN)$_{\rm faint}$ &  16.52     & 1036 & dE(di) &  12.94                        \\
1577 & dE(nN)$_{\rm bright}$ &  14.97    & 1235 & dE(nN)$_{\rm faint}$ &  16.75     & 1183 & dE(di) &  13.27                        \\
1647 & dE(nN)$_{\rm bright}$ &  15.14    & 1288 & dE(nN)$_{\rm faint}$ &  16.40     & 1204 & dE(di) &  15.35                        \\
1698 & dE(nN)$_{\rm bright}$ &  15.14    & 1298 & dE(nN)$_{\rm faint}$ &  16.97     & 1304 & dE(di) &  14.23                        \\
1704 & dE(nN)$_{\rm bright}$ &  15.57    & 1314 & dE(nN)$_{\rm faint}$ &  16.12     & 1392 & dE(di) &  13.84                        \\
1743 & dE(nN)$_{\rm bright}$ &  14.65    & 1337 & dE(nN)$_{\rm faint}$ &  16.66     & 1422 & dE(di) &  12.78                        \\
1762 & dE(nN)$_{\rm bright}$ &  15.67    & 1352 & dE(nN)$_{\rm faint}$ &  16.33     & 1444 & dE(di) &  15.04                        \\
1870 & dE(nN)$_{\rm bright}$ &  15.11    & 1370 & dE(nN)$_{\rm faint}$ &  16.57     & 1505 & dE(di) &  17.03                        \\
1890 & dE(nN)$_{\rm bright}$ &  14.06    & 1432 & dE(nN)$_{\rm faint}$ &  16.43     & 1514 & dE(di) &  14.40                        \\
1895 & dE(nN)$_{\rm bright}$ &  14.15    & 1438 & dE(nN)$_{\rm faint}$ &  16.69     & 1691 & dE(di) &  16.92                        \\
1948 & dE(nN)$_{\rm bright}$ &  14.78    & 1449 & dE(nN)$_{\rm faint}$ &  16.86     & 1695 & dE(di) &  13.46                        \\
1982 & dE(nN)$_{\rm bright}$ &  14.68    & 1464 & dE(nN)$_{\rm faint}$ &  16.73     & 1836 & dE(di) &  13.66                        \\
1995 & dE(nN)$_{\rm bright}$ &  14.96    & 1472 & dE(nN)$_{\rm faint}$ &  17.52     & 1896 & dE(di) &  14.07                        \\
2004 & dE(nN)$_{\rm bright}$ &  15.17    & 1482 & dE(nN)$_{\rm faint}$ &  17.07     & 1910 & dE(di) &  13.23                        \\
2008 & dE(nN)$_{\rm bright}$ &  14.04    & 1518 & dE(nN)$_{\rm faint}$ &  17.82     & 1921 & dE(di) &  14.37                        \\
2028 & dE(nN)$_{\rm bright}$ &  15.61    & 1543 & dE(nN)$_{\rm faint}$ &  16.99     & 1949 & dE(di) &  13.08                        \\
2056 & dE(nN)$_{\rm bright}$ &  15.33    & 1573 & dE(nN)$_{\rm faint}$ &  15.68     & 2019 & dE(di) &  13.56                        \\
2078 & dE(nN)$_{\rm bright}$ &  15.63    & 1599 & dE(nN)$_{\rm faint}$ &  16.68     & 2042 & dE(di) &  13.58                        \\
0011 & dE(nN)$_{\rm faint}$  &  16.15    & 1601 & dE(nN)$_{\rm faint}$ &  16.18     & 2048 & dE(di) &  12.99                        \\
0091 & dE(nN)$_{\rm faint}$  &  16.85    & 1622 & dE(nN)$_{\rm faint}$ &  16.82     & 2050 & dE(di) &  14.36                        \\
0106 & dE(nN)$_{\rm faint}$  &  17.11    & 1629 & dE(nN)$_{\rm faint}$ &  16.75     & 2080 & dE(di) &  14.96                        \\
0127 & dE(nN)$_{\rm faint}$  &  16.49    & 1650 & dE(nN)$_{\rm faint}$ &  16.32     & 0021 & dE(bc) &  14.08                        \\
0244 & dE(nN)$_{\rm faint}$  &  16.62    & 1651 & dE(nN)$_{\rm faint}$ &  16.33     & 0170 & dE(bc) &  13.50                        \\
0294 & dE(nN)$_{\rm faint}$  &  17.49    & 1652 & dE(nN)$_{\rm faint}$ &  16.07     & 0173 & dE(bc) &  14.21                        \\
0299 & dE(nN)$_{\rm faint}$  &  16.22    & 1657 & dE(nN)$_{\rm faint}$ &  16.40     & 0218 & dE(bc) &  14.00                        \\
0317 & dE(nN)$_{\rm faint}$  &  17.18    & 1658 & dE(nN)$_{\rm faint}$ &  15.82     & 0281 & dE(bc) &  14.77                        \\
0335 & dE(nN)$_{\rm faint}$  &  16.81    & 1663 & dE(nN)$_{\rm faint}$ &  16.56     & 0308 & dE(bc) &  13.14                        \\
0361 & dE(nN)$_{\rm faint}$  &  16.58    & 1682 & dE(nN)$_{\rm faint}$ &  16.24     & 0674 & dE(bc) &  16.18                        \\
0403 & dE(nN)$_{\rm faint}$  &  17.12    & 1688 & dE(nN)$_{\rm faint}$ &  15.93     & 0781 & dE(bc) &  13.90                        \\
0418 & dE(nN)$_{\rm faint}$  &  16.45    & 1689 & dE(nN)$_{\rm faint}$ &  16.44     & 0870 & dE(bc) &  14.10                        \\
0421 & dE(nN)$_{\rm faint}$  &  16.65    & 1702 & dE(nN)$_{\rm faint}$ &  16.51     & 0901 & dE(bc) &  16.49                        \\
0422 & dE(nN)$_{\rm faint}$  &  16.81    & 1717 & dE(nN)$_{\rm faint}$ &  15.80     & 0951 & dE(bc) &  13.35                        \\
0444 & dE(nN)$_{\rm faint}$  &  16.44    & 1719 & dE(nN)$_{\rm faint}$ &  17.55     & 1488 & dE(bc) &  14.12                        \\
0454 & dE(nN)$_{\rm faint}$  &  16.97    & 1729 & dE(nN)$_{\rm faint}$ &  17.76     & 1501 & dE(bc) &  14.90                        \\
0458 & dE(nN)$_{\rm faint}$  &  15.84    & 1733 & dE(nN)$_{\rm faint}$ &  16.59     & 1512 & dE(bc) &  14.82                        \\
0466 & dE(nN)$_{\rm faint}$  &  15.92    & 1740 & dE(nN)$_{\rm faint}$ &  16.64     & 1684 & dE(bc) &  14.45                        \\
0499 & dE(nN)$_{\rm faint}$  &  16.89    & 1745 & dE(nN)$_{\rm faint}$ &  16.28     & 1779 & dE(bc) &  14.01                        \\
0501 & dE(nN)$_{\rm faint}$  &  16.02    & 1764 & dE(nN)$_{\rm faint}$ &  15.96     & 1912 & dE(bc) &  13.26                        \\
0504 & dE(nN)$_{\rm faint}$  &  15.89    & 1792 & dE(nN)$_{\rm faint}$ &  16.99     &  & & 
  \enddata
\end{deluxetable}

%\clearpage

\clearpage
 
%\clearpage
 
\clearpage
 
\clearpage

%\clearpage
 
%\clearpage

\clearpage
 
\end{document}